;
\documentclass[%
 reprint,
 amsmath,amssymb,
 aps,
]{revtex4-2}

\usepackage{graphicx}
\usepackage{dcolumn}
\usepackage{bm}
\usepackage{booktabs}
\usepackage{subcaption} 

\begin{document}


\title{
Free energy calculations and unbiased dynamics reveal a continuous liquid-liquid transition in water no man's land}

\author{Alexandre Jedrecy}
\author{A. Marco Saitta}%
 \email{marco.saitta@sorbonne-universite.fr}
\author{Fabio Pietrucci}%
 \email{fabio.pietrucci@sorbonne-universite.fr}
\affiliation{%
Insitut de Min\'eralogie, de Physique des Mat\'eriaux et de Cosmochimie, Sorbonne Universit\'e, CNRS, MNHN, UMR 7590, Paris, France
}%


\date{\today}

\begin{abstract}
The existence of a first-order phase transition between a low-density liquid (LDL) and a high-density liquid (HDL) forms of supercooled water has been a central and highly debated issue of physics and chemistry in the last three decades. We present a computational study that allows to determine the free-energy landscapes of supercooled water over a wide range of pressure and temperature conditions, using the accurate TIP4P/2005 force field. Our approach combines topology-based structural transformation coordinates, state-of-the-art free-energy calculation methods, and extensive unbiased molecular dynamics simulations. All our results indicate that transitions between the LDL and HDL forms are smooth throughout the so-called "no man's land", and that free-energy barriers do not appear until the onset of the solid, non-diffusive amorphous forms.
\end{abstract}

\maketitle


\section{\label{sec:intro} Introduction}

Among the many peculiarities and anomalies of water, by far and large the most important and common chemical substance on the surface of the Earth and in living organisms, its polyamorphism at low temperatures has clearly been one of the most puzzling in the last three decades. The experimental evidence of three amorphous forms~\cite{Mishima-nat-84, Mishima-nat-85,Finney-prl-2002}, and of apparently reversible first-order transitions among some of them~\cite{Klotz-prl-05}, seems in fact at odds with the very thermodynamic notion of metastable glassy forms. 
Several scenarios have been formulated to explain these phenomena, the most famous, and somehow controversial, being the occurrence of a first-order  liquid-liquid transition in supercooled water, extending at lower temperatures in the amorphous region, and terminating with a second critical point at higher temperatures. This hypothesis, formulated on the basis of a computational molecular dynamics study, based on a model potential for water~\cite{Poole-n-92}, is however extremely challenging to be verified by experiments in pure bulk water. Indeed, its supposed location would lie below the kinetic limit of homogeneous ice formation, in the so called {\it no man's land}~\cite{caupin-jncs-15, Amann-rmp-16, Gallo-cr-16, Palmer-cr-18}.

Important theoretical efforts have since been made, in order to improve our understanding of polyamorphism, notably in the liquid phase. For example, the definition of a two-states model provides a unitary description of the thermodynamics for most polymorphic fluids~\cite{Anisimov-prx-18}. In this view, water is considered as a ``mixture'' of two interconvertible local structures: a high-density, high-entropy liquid and a low-density, low-entropy liquid~\cite{tanaka-jcp-00, Holten-sr-12, Russo-ncom-14, Esposito-pre-06}. The model predicts four scenarios, discriminated through the density extrema loci~\cite{Anisimov-prx-18}:
\begin{itemize}
   \setlength\itemsep{0.005cm}
    \item a singularity-free scenario, with interconversion between two states but no phase separation;
    \item a liquid-liquid critical point scenario, with interconversion and phase separation;
    \item a degenerate case where the critical point coincides with the vapor-liquid spinodal;
    \item a critical-point-free scenario, with a virtual critical point located below the vapor-liquid spinodal.
\end{itemize}
This model is elegant, and its predictions intriguing, but critically dependent on the detailed choice of the thermodynamic parameters.
A recent work explicitly demonstrates the above statement and settles some long-standing issues about two-state models using a mean-field approach on a two-component compressible lattice model allowing interconversion of the two components, able to reproduce the anomalies of water \cite{Caupin21}. Details of the interactions determine the existence or absence of a discontinuous transition and of a critical point, while the often relied-upon proxies result misleading, since, citing the Authors, "for a given fluid, neither the shape of the line of density maxima, nor that of the liquid spinodal limit, nor the existence of $\kappa_T$ or $C_P$ maxima, is sufficient to identify which scenario is valid" \cite{Caupin21}. One logical consequence, from the computational viewpoint, is that interatomic potentials customarily employed to perform molecular dynamics simulations of water could produce different scenarios depending on slight differences of formulation ({\sl vide infra}).

From the experimental point of view, a huge battery of diverse set-ups have been deployed, over the years, using for example aqueous solutions, or confined/micro-sized systems, in order to overcome the thermodynamic frontiers of the no man's land, while avoiding the inevitable crystallization of supercooled water into ice~\cite{caupin-jncs-15}. Some of those experiments have been able to firmly establish that the transition between the low and high density amorphous ices is a first order one~\cite{Klotz-prl-05, Nelmes-np-06, Saitta-jpcb-06, Perakis-pnas-17, Handle-jcp-18-exp, Shen-prm-19}. Some experiments were able to give  hints on the presence of two competing liquid forms in the supercooled region~\cite{Nilsson-cp-11, Singh-pnas-17, Perakis-pnas-17, Lin-pnas-18, Pathak-jcp-19}. For bulk water, recent experiments carried out at negative pressures suggest that the right scenarios are either the singularity free or the liquid-liquid critical point~\cite{Pallares-pnas-14}. Other experiments conducted on salty water, give instead strong arguments against a first order liquid-liquid transition in the supercooled region~\cite{Bove-jcp-19}, even if a first order transition was observed for the corresponding amorphous phases~\cite{Bove-prl-11}, hence showing that no direct link necessarily exists between polyamorphism and a liquid-liquid transition.

From the computational point of view,
the second critical point scenario has been a long source of debate since its very first proposition~\cite{Poole-n-92}, mostly because that work was based on the "ST2 model" of water~\cite{Stillinger-jcp-74}, which is known to be significantly overstructured, and thus to "enhance" certain anomalies of water. After several free-energy studies found contradicting results with this model, either demonstrating the LDL-HDL transition and coexistence~\cite{Harrington-prl-97, liu-jcp-2012, Poole-jcp-13}, in systems containing up to 600 ST2 water molecules~\cite{Palmer-n-14}, or supporting a no-transition scenario with up to 512 molecules~\cite{Limmer-jcp-11, Limmer-jcp-13}, a consensus emerged on the former hypothesis, thus validating phase coexistence and reconciling the two independent free energy calculations~\cite{Palmer-jcp-18, Palmer-cr-18}. However, this result seems limited to this specific model, nowadays known for its drawbacks, and widely considered as not particularly representative of real water.
Other studies pointed out in fact that the thermodynamics of the putative LDL-HDL transition in supercooled water was heavily model-dependent~\cite{Brovchenko-jcp-05,Huggins-jcp-12,Holten13}. In the last few years, the so-called TIP4P/2005 force field~\cite{Abascal-jcp-05} has emerged as one of the most accurate models, as it reproduces quite accurately the phase diagram and anomalies of water~\cite{Corradini-jcp-10, vega-faraday-09, Conde-jcp-13, Conde-jcp-17}. Several numerical studies were performed with TIP4P/2005 to assess the existence of a liquid-liquid transition, although none with a thorough and extensive free-energy approach.
A critical point for TIP4P/2005 water was first proposed at 1.35 kbar, 193 K and 1012 kg.$\mathrm m^{-3}$, based on the analysis of density and concentration fluctuations in the supercooled region in 500-molecules models for durations of 500 ns~\cite{Abascal-jcp-10}. A subsequent analysis failed to reproduce this result with larger boxes and longer simulations (1,000 to 32,000 molecules and 500 ns to 5 $\mu$s) and showed that size effects are important, together with the long relaxation time of the system~\cite{Overduin-jcp-13}, as it was confirmed later on~\cite{Limmer-mp-15, Palmer-mp-16}. Another study looked at density fluctuations concluding that they constitute the signature of a liquid-liquid transition~\cite{Yagasaki-pre-14}, but once again a subsequent analysis with larger simulation boxes argued that their origin is the appearance of ice-like structure~\cite{Overduin-jcp-15}. With the coupled use of longer simulations and a two-state thermodynamic analysis, a new critical point was proposed at 182 K and 1.70 kbar ~\cite{Singh-jcp-16, Biddle-jcp-17}, consistently with previous numerical~\cite{Overduin-jcp-13} and experimental studies~\cite{Pallares-pnas-14}. 

Two recent computational studies further revised the predicted locations of the critical point. In the first study ~\cite{Shi20}, a two-state thermodynamic model was fit to data from TIP4P/2005 simulations (1000 molecules) above 182 K, and extrapolation of the static and dynamic Schottky lines to lower temperature was interpreted as a prediction of a critical point at 172 K and 2.16 kbar, while in the case of TIP5P, the predicted location is 216 K and 2.58 kbar.
In the second study ~\cite{debenedetti-science-20}, based on the analysis of density fluctuations, the authors combined extensive unbiased simulations of tens of $\mu$s for 300, 500 and 1000 molecules, at $T\ge 177$ K for TIP4P/2005 and $\ge 188$ K for TIP4P/ice, i.e., above the postulated second critical point of the two models (see below), with an histogram reweighting technique to extrapolate order parameter distributions at lower temperature, closer to the supposed critical regime. By fitting the extrapolated distributions, together with static scattering functions computed on larger boxes at $T>180$ K, to a 3D Ising model, an estimation of the liquid-liquid critical point conditions is obtained at $T_c = 172 \pm 1$ K and $P_c = 1861 \pm 9$ bar for TIP4P/2005. This elegant work still is not a proof of the liquid-liquid phase transition (LLPT), as rigorous proof requires performing free energy calculations at subcritical temperatures~\cite{debenedetti-science-20}.

The present work aims at providing a robust answer to this long-going question by a combination of free energy calculations and unbiased simulations in no man's land. To this end, we adopt a strategy based on several methodological strengths. 
First, we employ a versatile topological metric to describe structural transformations in water, already proved to be very effective in discriminating the known crystalline, amorphous, and liquid forms of water~\cite{Pietrucci-jcp-15}, and that we successfully used to study several phase transitions throughout the phase diagram of water, including the extremely challenging spontaneous nucleation of crystalline ice from the bulk liquid~\cite{PIV-silvio}.
Second, we exploit a synergistic free-energy calculation approach, combining metadynamics to explore the configuration space, umbrella-sampling to collect extensive statistics along the transformation paths, 
and unbiased MD trajectories probing the spontaneous evolution from different phase-space regions to validate free energies and extract valuable dynamic information. 
Third, we use the TIP4P/2005 force field, often considered the most reliable and accurate to describe real water. We fully describe our approach in the Materials and Methods section.
Anticipating our results, the combination of these advanced techniques and demanding calculations allows us to establish the relative "flatness" of the free-energy landscapes throughout the no man's land, and thus to suggest that no LDL-HDL first-order transition exists in the supercooled regime, differently from what is experimentally observed in the amorphous region.

\section{\label{sec:results} Results}

One of the main results of our work is the systematic accurate calculation of free-energy landscapes at different $P,T$ conditions corresponding to the putative liquid-liquid transition.
As order parameter, we employed the $S$ path coordinate introduced in Ref.~\cite{PIV-silvio}: the definition takes into account the network of interatomic connections in the first and second neighbor shells (as encoded in permutation invariant vectors~\cite{Gallet13}), and the progress of the transformation is calculated by comparing the topology of the MD configuration with those of reference LDL and HDL structures (see section Materials and Methods for technical details).

\begin{figure*}[!ht]
     \centering
    \includegraphics[width=0.95\linewidth]{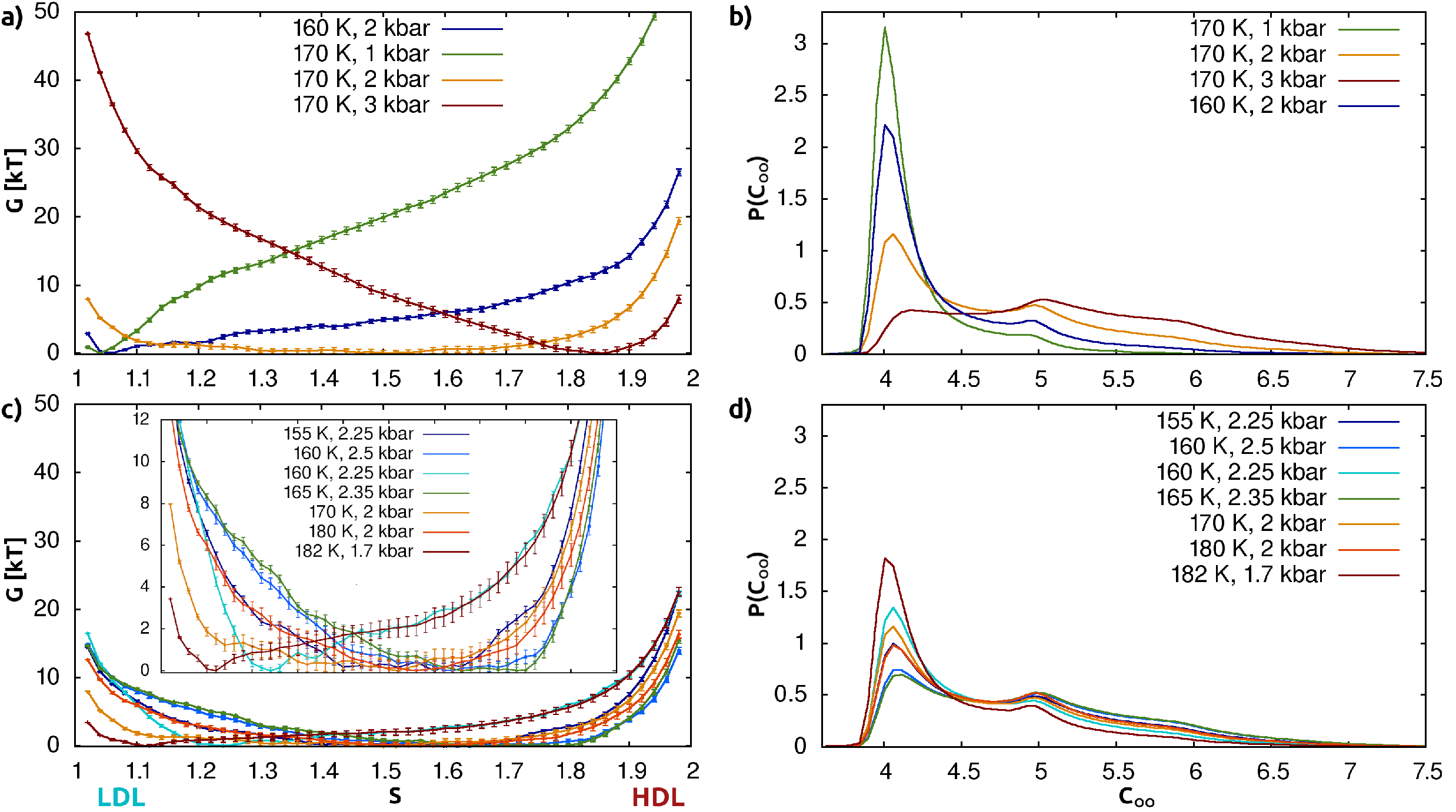}
    \caption{
    (a,c) Free-energy profiles for the low-density/high-density water transformation, and (b,d) corresponding distributions of oxygen-oxygen coordination numbers ($C_{OO}$, see definition in Materials and Methods). $S\approx1.1$ correspond to low density and $S\approx1.9$ to high density.
    (a,b) Pressures and temperature around 170 K and 2 kbar. 
    (c,d) Conditions intermediate between those favoring low density and high density; note the relatively flat free-energy profiles (see the zoomed inset) and broad $C_{OO}$ distributions. 
    }
    \label{fig:FES}
\end{figure*}

\begin{figure}[!ht]
    \centering
    \includegraphics[width=0.95\linewidth]{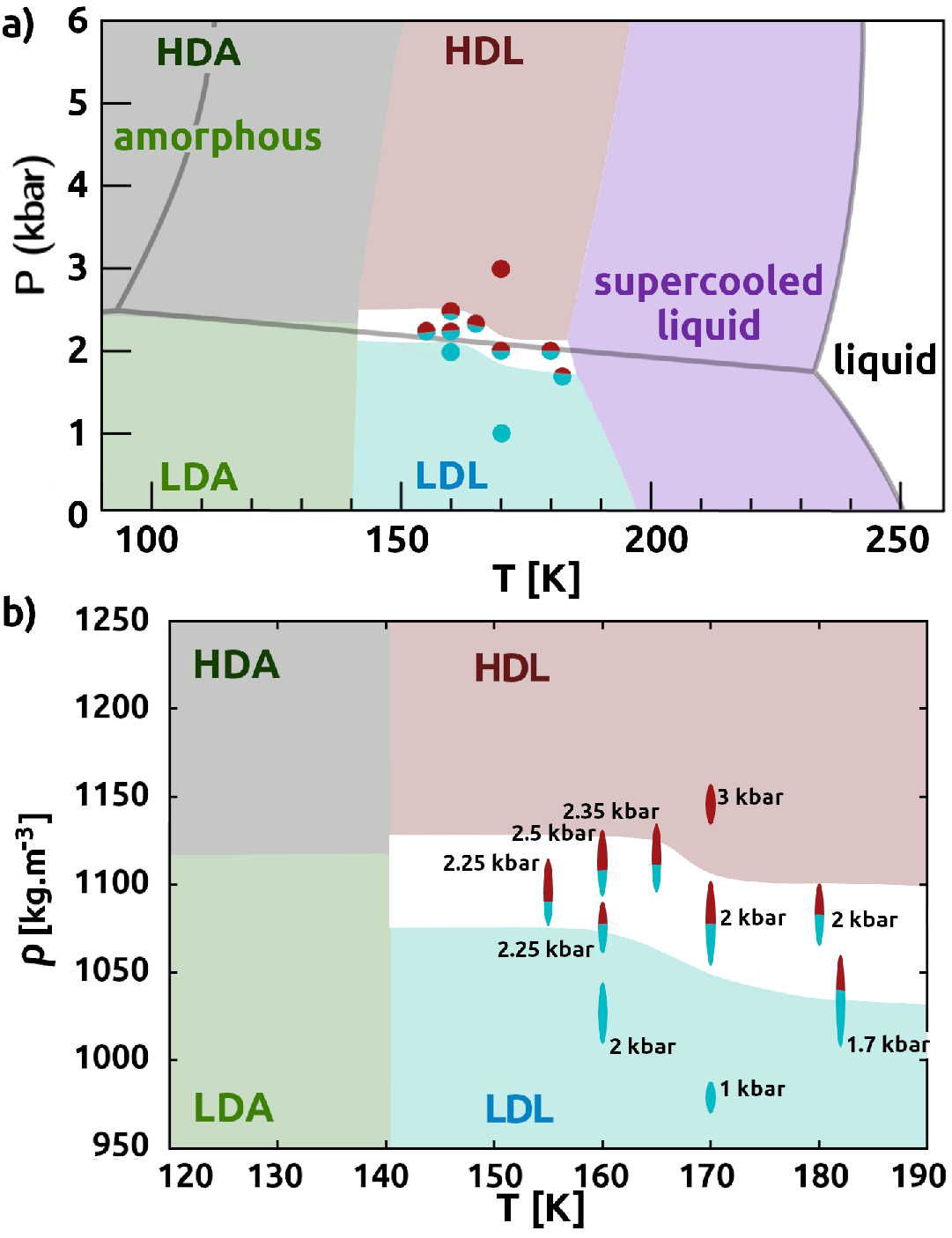}
    \caption{
    (\ref{fig:phase_diagram}.a) Schematic phase diagram of TIP4P/2005 water in the $P,T$ region considered in this work. The solid gray lines indicates the stable phases~\cite{Conde-jcp-13}. Dots represent the conditions of MD simulations. Blue and red dots correspond to LDL and HDL, respectively. Dots half-red and half-blue indicate a nearly flat free-energy profile spanning low- to high-density water. The same color scheme is adopted to indicate areas where each of the two forms is expected to prevail. In the white areas the system is neither clearly LDL nor HDL.
    Low- and high-density amorphous forms are indicated in light- and dark-green colors, respectively. 
    (\ref{fig:phase_diagram}.b) Density as a function of temperature for the same phase diagram. The average density and its standard deviation (height of the ellipsoids) are computed by re-weighting the density values in umbrella sampling simulations with the equilibrium population $\mathrm{e}^{-G(S)/k_BT}$, as a functon of the $S$ path coordinate.}
    \label{fig:phase_diagram}
\end{figure}

Figure~\ref{fig:FES} presents the free energy profiles at different $P,T$ conditions, and the corresponding equilibrium distribution of oxygen-oxygen coordination numbers $C_{OO}$. 
As a guide to interpret the graphs, low-density water features $S\lesssim1.5$, $C_{OO}\lesssim 4.5$, and the opposite for high-density water. We remark the high correlation between $S$ and density in supercooled water, being almost proportional to each other (see SI Fig. 2). 
Most interestingly, by following the 170 K isotherm in Fig.\ref{fig:FES}.a or the 
approximate isobar at $2-2.5$ kbar
 in Fig.~\ref{fig:FES}.c the free-energy profiles always exhibit a single minimum along the transformation path.

For fixed $T$ and increasing $P$, the most probable form of the system progressively switches from low-density water to high-density water, as displayed by the free-energy minimum moving from low to high $S$ values in Figure~\ref{fig:FES}.a. 
A natural question arises: what is the precise shape of the free-energy landscape at conditions where low- and high-density water forms are equiprobable?

Remarkably, for $P,T$ conditions that are intermediate with respect to those favoring low density or high density it is possible to observe relatively flat free energy profiles (within a few $k_BT$ units), without any sizable barrier separating LDL and HDL (Figure~\ref{fig:FES}.c).
Such flat profiles indicate that the system populates a relatively broad range of different densities and coordination numbers, as can indeed be observed in Fig.~\ref{fig:FES},~\ref{fig:phase_diagram},  and in SI Fig. 12. 

Figure~\ref{fig:phase_diagram}.a summarizes these results in a schematic phase diagram that we reconstruct in no man's land, based on the structural 
and dynamical 
features of the low-free-energy part of configuration space, within 2 $k_BT$ units from the minimum.
We distinguish liquid and amorphous phases based on the diffusion coefficient, following a previous study on TIP4P/2005 water~\cite{Wong-jcp-15} (see also SI Table 3).
Average water densities and relative fluctuations at each $P,T$ point are reported in Figure~\ref{fig:phase_diagram}.b (see also SI Fig. 12 for more extensive data). %

\begin{figure}[!ht]
    \centering
    \includegraphics[width=0.95\linewidth]{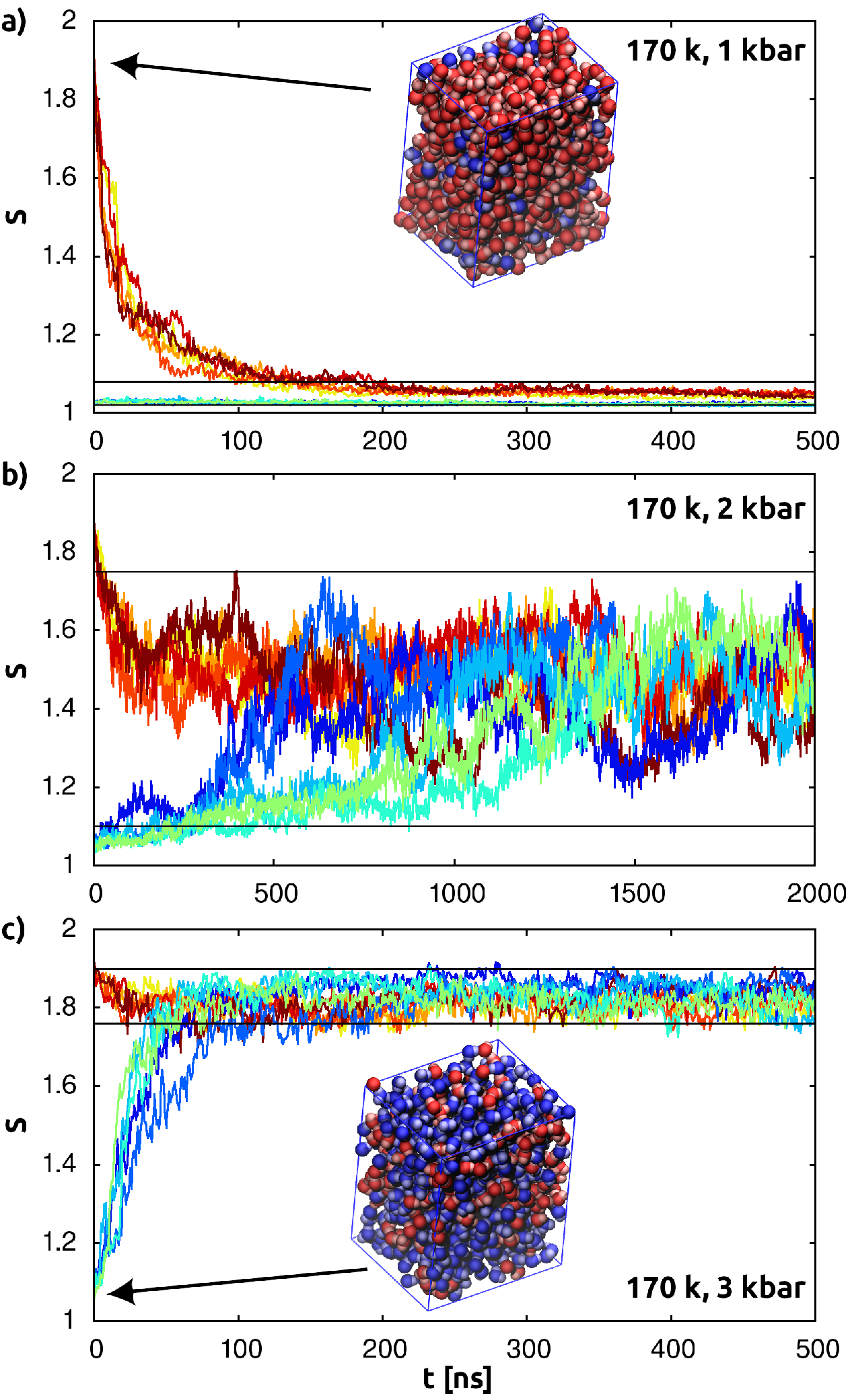}
    \caption{
        Independent unbiased MD trajectories initiated from LDL (blue) or HDL (red) umbrella sampling end-point configurations, at 170 K and three different pressures (note the different horizontal scales). The intervals delimited by black lines correspond to free-energy values within 2 $k_BT$ from the minimum as reconstructed from umbrella sampling (Figure~\ref{fig:FES}.a).
        (a) At 1 kbar, the initial high-density configuration relaxes within $\approx100$ ns to the free-energy minimum of low density; an initial low-density configuration retains its character. 
        (b) At 2 kbar, both the initial high-density and low-density configurations relax within the very broad free-energy basin of intermediate density, displaying slow diffusion on the timescale of hundreds of ns. 
        (c) At 3 kbar, the initial low-density configuration relaxes within $\approx100$ ns into the free-energy minimum of high density; an initial high-density configuration retains its character.
    }
    \label{fig:unbiased}
\end{figure}

At this point, the natural question becomes: do the large density fluctuations in the white band of Fig.~\ref{fig:phase_diagram} correspond to coexistence of two distinct water forms, low-density and high-density, and hence two metastable states?  
To address this relevant issue we generated tens of long {\it free and unbiased} MD trajectories, with a cumulative duration of more than 145 microseconds (a small selection being shown in Figure~\ref{fig:unbiased}, see also SI Fig. 7). 
We anticipate that such trajectories allow also to validate the shape of the free-energy profiles reconstructed with umbrella sampling.

Figure~\ref{fig:unbiased} shows unbiased trajectories at 170 K initiated from 
the end-point of
low- or high-density umbrella sampling simulations. Comparison with Figure~\ref{fig:FES}.a demonstrates that MD trajectories behave as expected from the computed free-energy profiles, relaxing from high- towards low free-energy regions according to the slope of the profile (i.e., the mean force), until showing stationary free diffusion in the region of the minimum. The latter is well-localized at low density at 1 kbar, it has a broad shape at 2 kbar, and is well-localized at high density at 3 kbar, as discussed above. 
As a further quantitative benchmark, the density distributions reconstructed from unbiased trajectories are in good agreement with those reconstructed from the equilibrium free energy profiles obtained by umbrella sampling (SI Fig. 10 and 11).
Hence, unbiased MD is consistent with enhanced sampling simulations and it represents an independent robust validation of the reconstructed free-energy landscapes. Once again, we never observe local kinetic trapping of the system in two distinct states: at all $P,T$ conditions and irrespective of the starting density the system steadily relaxes towards a single precise region in configuration space, without evident bottlenecks.

\begin{figure}[!ht]
    \centering
    \includegraphics[width=0.95\linewidth]{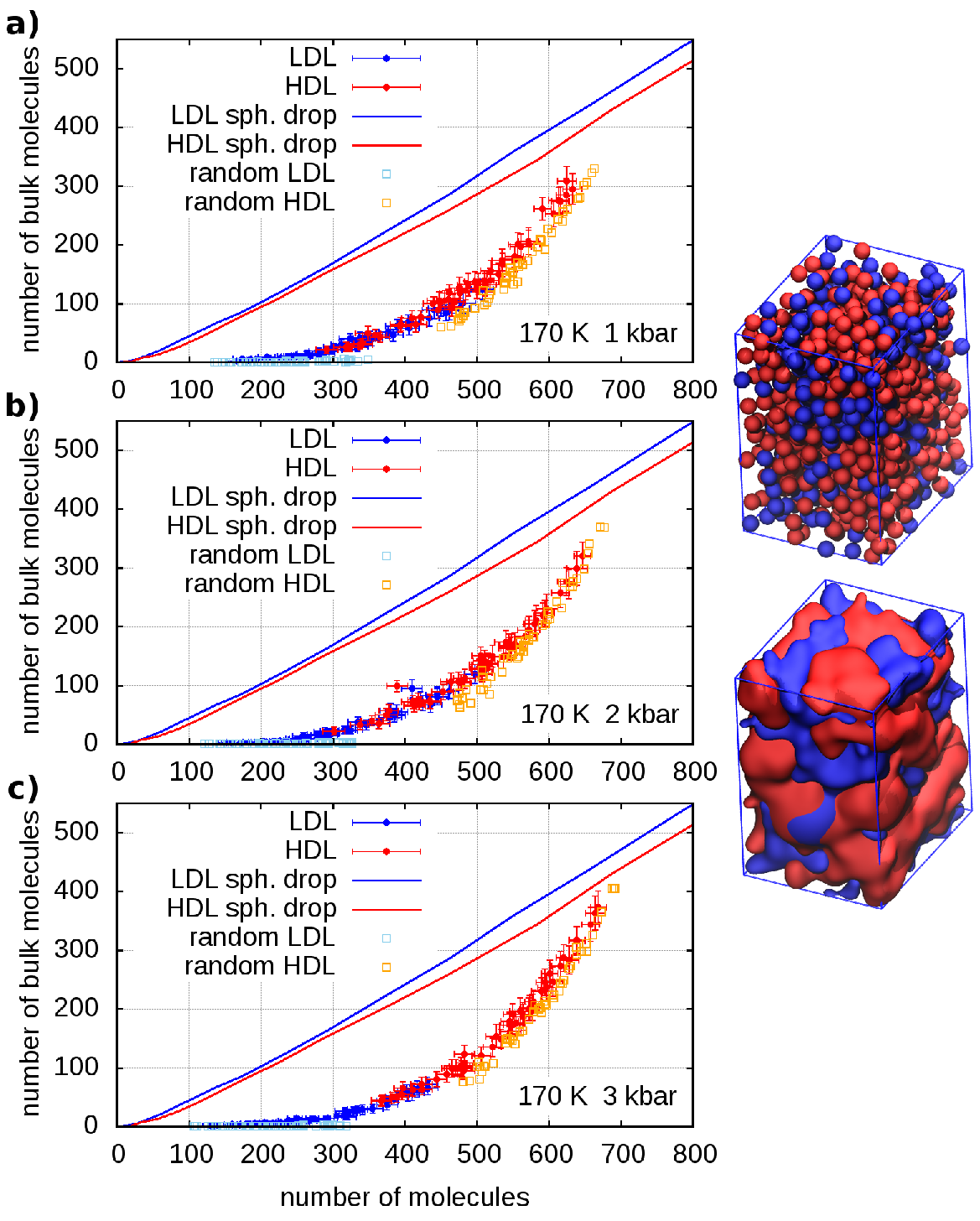}
    \caption{
    Number of LDL-like and HDL-like molecules (assigned on the basis of $C_{OO}<4.5$ or $>4.5$, respectively) that are surrounded by molecules of the same type, {\it i.e.} not at the interface LDL/HDL, extracted from umbrella sampling trajectories. For comparison, continuous curves indicates the number of bulk molecules in a spherical droplet containing only LDL or HDL, and the squares correspond to random networks of molecules with the same bond distribution as LDL or HDL in MD configurations (see section Materials and Methods for details). The 3D structure (balls and surfaces enclosing them) illustrate a typical configuration at 170 K and 2 kbar.
    }
    \label{fig:bulkfraction}
\end{figure}

As a final benchmark, we analyzed the structure of instantaneous atomic configurations, with particular attention to $P,T$ conditions maximizing density fluctuations, to understand whether low- and high-coordinated water molecules are randomly mixed or they group together in order to minimize the LDL/HDL interface. Clearly, the hypothesis of a coexistence of two {\sl distinct} liquid forms requires the existence of a well-defined geometrical interface characterized by unfavorable molecular interactions, hence of minimal extension (spherical or planar). Under such hypothesis, as in classical nucleation theory, the interface provides an unfavorable free-energy contribution to the total budget of the system, creating a barrier that grows with system size as $N^{2/3}$ (as observed in ST2 water in Ref.~\cite{Palmer-n-14}).

Visual inspection both of unbiased MD trajectories and of umbrella sampling trajectories does not reveal a clear tendency towards separation of large and convex LDL or HDL regions: the respective clusters of hydrogen-bonded molecules display a complex, interpenetrating interface whose extension appears far from minimal.
A quantitative assessment is presented in Figure~\ref{fig:bulkfraction}: molecules are identified as LDL-like or HDL-like based on $C_{OO}<4.5$ or $>4.5$, respectively, and for each type the number of bulk molecules (i.e., in contact only with alike molecules, thus not at the interface) is plotted against the total number. In principle, the fraction of bulk molecules is maximized when all molecules of one type form a single spherical drop (or a flat periodic slab), and it is minimized when molecules are randomly mixed. These two limits are also represented in Figure~\ref{fig:bulkfraction} (see also SI Fig.  8 and 9), allowing to appreciate how MD configurations at putative coexistence conditions (i.e., with similar LDL and HDL fraction) are in reality much closer to a randomly intermixed system than to one exhibiting phase-separation. We find similar results at all $P,T$ conditions explored, both for umbrella sampling and unbiased trajectories, whenever both LDL and HDL are present in significant amount. 
Previous studies addressed the number and size of LDL/HDL clusters in TIP4P/2005 water, albeit at $T\ge 190$ K and without discussing the interface shape~\cite{Martelli19}. In summary, our structural analysis is once again consistent with the absence of liquid-liquid phase separation.

\section{\label{sec:discussion} Discussion}

Our results show that the metric based on permutation invariant vectors~\cite{Gallet13,PIV-silvio} resolves well the range of supercooled water structures and densities throughout the vast $P,T$ region explored. This result extends the analyses in Ref.~\cite{Pietrucci-jcp-15}, where the same metric was demonstrated able to resolve and clusterize  structures belonging to liquid, amorphous and crystalline water. In combination with path coordinates, the metric allowed here also to reconstruct free energy landscapes extending the approach of Ref.~\cite{PIV-silvio}, applied also to heterogeneous ice nucleation in Ref.~\cite{Fitzner17}, to transitions between supercooled liquid forms. While several other order parameters have been applied to specific investigations on water~\cite{Gallo-cr-16,Holten13,Niu19},
due to its generality our computational approach allowed a comprehensive and unitary study of water structure, dynamics and thermodynamics encompassing liquid polymorphs, solid polyamorphs and crystals.

We performed enhanced sampling and unbiased MD simulations in a range of $P,T$ conditions between 155 and 182 K and between 1 and 3 kbar, and in all cases we could not find any compelling evidence of liquid-liquid phase separation and of a corresponding free-energy barrier.
We reach this conclusion employing three different and complementary methods: 1) enhanced sampling simulations to reconstruct free-energy landscapes for the low- to high-density transition, 2) long unbiased MD simulations to probe the putative local stability of LDL and HDL phases and to confirm free-energy landscapes, and 3) in-depth structural analysis of clusters formed by low- and high-density water to assess the geometric properties of the LDL/HDL interface, a crucial indicator of phase separation.

In particular, 
at $P,T$ conditions close to the most recently predicted locations of 
the liquid-liquid critical point (182 K, 1.7 kbar and 180 K, 2 kbar to compare with Ref.~\cite{Singh-jcp-16}, and 170 K, 2 kbar to compare with Ref.~\cite{debenedetti-science-20})
we found no free-energy barrier and a single broad minimum (Fig.~\ref{fig:FES}.c), characterized by significant fluctuations in density and coordination number (Fig.~\ref{fig:phase_diagram}.b,~\ref{fig:FES}.d), without evidence of phase separation between LDL and HDL. These qualitative features, however, are not unique of a single point in $P,T$-space, since we could follow a line of points with similar behavior -- in particular without free-energy barrier -- from 182 K down to at least 155 K, close to the frontier with amorphous water. 

The situation is different for the ST2 model: in Ref.~\cite{Palmer-n-14} a 4 $k_BT$ free-energy barrier separating LDL from HDL could be measured at 229 K and 2.4 kbar for a system size of 192 molecules, based on extensive and careful enhanced sampling simulations; the barrier was confirmed by unbiased Monte Carlo simulations reversibly sampling the LDL-HDL transition, and it was shown to scale like $N^{2/3}$ for $198\le N\le 600$, as expected for a first-order phase transition. We remark that we could not find any barrier with the more accurate TIP4P/2005 force field despite a system size of 800 molecules, larger than the largest one considered in Ref.~\cite{Palmer-n-14}.

From the viewpoint of the physics of supercooled water, the most important lesson delivered by freely relaxing trajectories is that there is no $P,T$ point (within the broad range we explored) where LDL and HDL are both kinetically trapped in their respective forms for a measurable time (Fig.~\ref{fig:unbiased} 
and SI Fig. 7). We must conclude that it is impossible to observe LDL and HDL as distinct and persistent forms at the same thermodynamic conditions. In other words, coexistence of the two phases is impossible, so that LDL and HDL are not two distinct phases in the thermodynamic sense. We remark that while the existence of a mechanically stable LDL/HDL interface has been demonstrated for the ST2 model up to large system sizes~\cite{Yagasaki-pre-14,Guo-mp-18},
such a demonstration is lacking for the more reliable and accurate TIP4P force fields family \cite{Palmer-cr-18}.

On the contrary, we conclude that it is possible to change the form of water from lower-density and lower-coordination values to higher ones in a continuous way, for instance by increasing pressure from $\approx$1.5 to $\approx$3 kbar at any temperature between 155 K and 182 K (Fig.~\ref{fig:phase_diagram}), without encountering a bottleneck in phase space, i.e., a barrier. Of course the timescale necessary for the system to relax from an initial out-of-equilibrium density slows down when lowering $T$, from $\approx$500 ns at 170 K and $1-3$ kbar (Fig.~\ref{fig:unbiased}) to 
 $\approx$2 $\mu$s at 160 K and $2.25-2.5$ kbar (see SI Fig. 7), 
however such a slow evolution appears the result of continuous diffusion 
in density space
with a weak diffusion coefficient, rather than of Poisson-distributed rare jumps across a barrier. This factual observation of the behavior of unbiased MD trajectories is fully consistent with our enhanced sampling simulations, where no free-energy barrier could be measured, and also with our analysis of the three-dimensional structure of low- and high-density water clusters, that revealed no strong tendency to minimize the interface area and a situation closer to random intermixing of LDL-like and HDL-like molecules than to phase separation (Fig.~\ref{fig:bulkfraction}).

Our observation of the lack of kinetic trapping is not compatible -- at least for the system size we considered -- with hypotheses evoked in the literature on the existence of an unfavorable interfacial free energy preventing the formation of two liquid phases in finite-sized systems, or on a phase-separation dynamics much slower than simulation times of the order of hundreds of nanoseconds~\cite{Yagasaki-pre-14,Overduin-jcp-15,Singh-jcp-16}.
All our results indicate the lack of a first-order liquid-liquid phase transition and of the related critical point for the accurate TIP4P/2005 water force field, thus leading to discard the scenarios that include such features and that have been hypothesized in the last 40 years to explain water anomalies~\cite{Gallo-cr-16,Palmer-cr-18}. This conclusion underlines the difficulty in extrapolating observations from other regions of the phase diagram deep into no man's land, and it is fully compatible with the most up-to-date statistical-mechanical analyses of two-state models able to reproduce water anomalies, that demonstrated how such anomalies are by no means a strong indication of a discontinuous transition and a second critical point, the latter being instead present or absent depending from the details of the interaction forces \cite{Caupin21}. Due to the rich and peculiar phenomenology of water physics and chemistry, the inescapable primary sources of information remain today experiments and atom-detailed computer simulations that directly probe the $P,T$ conditions of interest.




\section*{\label{sec:simu-methods} Simulation methods}

\subsection*{Molecular dynamics}

We performed molecular dynamics simulations employing the TIP4P/2005~\cite{Abascal-jcp-05} interatomic potential, in periodically-repeated triclinic boxes 
(see SI Fig. 8 and SI Tables 1, 2)  
containing $N = 800$ water molecules.
All simulation were done under $NPT$ conditions between $155-182$ K and $1-3$ kbar, employing the GROMACS 5.1.4 simulation package~\cite{gromacs}, for a total of 
$\sim 2.5\;10^6$ CPU-hours. 
We adopted a 2 fs timestep. Short-range interactions were truncated at 0.85 nm and the particle mesh Ewald method was used to compute electrostatic interactions. Bond constraints were maintained using the LINCS algorithm.
To control the temperature we used the stochastic velocity rescaling thermostat with a relaxation time of 0.5  ps~\cite{v-rescale-Bussi}, 
for the pressure we used an isotropic Parinello-Rahman barostat with a relaxation time of 2 ps~\cite{Parinello-jap-81}.

To generate the starting states of our simulation, we generated and equilibrated a liquid box at 180 K and 1 bar for 10 ns. Then we cooled down the box by step of 10 K for 5 ns to get temperatures ranging from 140 K to 180 K. For each temperature we performed a compression, increasing the pressure by step of 0.5 kbar for 5 ns until reaching 5 kbar. Note that here we are not achieving structural equilibration, however these states are just the starting points for metadynamics, that in turn is followed by umbrella sampling for harvesting statistics.

We performed unbiased MD starting from 
end-point
umbrella sampling  configurations of type LDL and HDL (see below), to observe the spontaneous relaxation of the system and the coherence with respect to umbrella sampling free-energy landscapes. 
We generated the following trajectories:
15$\times$5,000 ns at 160 K, 2.5 kbar; 7$\times$4,000 ns at 160 K, 2.25 kbar; 10$\times$500 ns at 170 K, 1 kbar; 10$\times$2,000 ns at 170 K, 2 kbar; 10$\times$500 ns at 170 K, 3 kbar; 10$\times$1500 ns at 180 K, 2 kbar.

\subsection*{Order parameter}

As order parameters able to distinguish low and high-density liquid configurations, in this work we adopted path collective variables \cite{path-CV} based on Euclidean distances of permutation invariant vectors (PIV) as a metric~\cite{Gallet13,Pietrucci-jcp-15,PIV-silvio}. We define the PIV 
starting from the following matrices built from Cartesian positions of oxygen and hydrogen atoms:
\begin{align*}
    v_{ij}^{OO} &= 1\times \mathsf{F} \left( \sqrt[\leftroot{-1}\uproot{3}\scriptstyle 3]{\frac{V}{V_0}} |\mathbf{r}_i^O - \mathbf{r}_j^O| \right) \\
    v_{ij}^{HH} &= 0.2\times \mathsf{F} \left( \sqrt[\leftroot{-1}\uproot{3}\scriptstyle 3]{\frac{V}{V_0}} |\mathbf{r}_i^H - \mathbf{r}_j^H| \right)
\end{align*}
where we used the switching function $\mathsf{F} = (1 - (r/r_0)^4) / (1 - (r/r_0)^{10})$, with $r_0 = 0.5$ nm, and reference volume $V_0 = 0.024 \; \mathrm{nm}^3$. 
In each matrix, we keep only elements with $i>j$ and sort them in ascending order to enforce invariance under permutation of identical atoms. Finally, we concatenate the two resulting vectors to form a PIV $\mathbf{V}$.

To define the path collective variables we used as metric the squared Euclidean distance between PIVs of atomic configurations ($D_{AB} = |\mathbf{V}_A - \mathbf{V}_B |^2$), such that with two reference states for a configuration $X$ we have
\begin{align*}
    S (X) &= \frac{1\times\mathrm{e}^{-\lambda D_{1X}} + 2\times\mathrm{e}^{-\lambda D_{2X}}} {\mathrm{e}^{-\lambda D_{1X}} + \mathrm{e}^{-\lambda D_{2X}}} \\
    Z (X) &= -\frac{1}{\lambda} \mathrm{log} \left( \mathrm{e}^{-\lambda D_{1X}} + \mathrm{e}^{-\lambda D_{2X}} \right)
\end{align*}

The two reference states correspond here to the equilibrated configurations at low- (1 bar) and high pressure (5 kbar) for every temperature, with $\lambda = 0.3$ set from $\lambda D_{12} \approx 2.3$, in order to obtain a smooth landscape \cite{PIV-silvio}.
The switching function $\mathsf{F}$ in the PIV definition is designed to include the first and second neighbor peaks in the $g(r)$ of our reference states (see SI Fig. 1). 

The order parameter $S$ adopted in this work is very general, as it can be applied to transitions between ordered or disordered structures in different materials \cite{PIV-silvio}; in the specific case of the LDL-HDL transition $S$ is highly correlated with the density of the system (see SI Fig. 2). 

\subsection*{Free energy calculations}

At selected $P,T$ conditions we performed enhanced-sampling simulations aimed at reconstructing the free-energy landscape for the supercooled liquid, using the open-source, community-developed PLUMED library version 2.6~\cite{Plumed}. First, we exploited metadynamics to obtain transition pathways as well as a preliminary estimate of the free energy landscapes~\cite{Laio-pnas-02}. Next, we reconstructed statistically converged free energy profiles with more expensive umbrella sampling simulations~\cite{Torrie-jcp-77}.

For metadynamics, we have done simulations of 25 ns to 50 ns, placing gaussian hills with $\sigma_s = 0.015$, $\sigma_z = 0.15$ and height of 1 kJ every ns (see SI Fig. 5).
To efficiently reconstruct free-energy profiles we exploited umbrella sampling simulations initiated from configurations explored with metadynamics. During metadynamic exploration the $Z$ variable did not provide important additional information with respect to $S$, so we have only enhanced the sampling of the latter. At each $P,T$ point, sampling was done with 48 windows spaced by $\delta_S = 0.02$, ranging from $1.02$ to $1.98$ in the $S$ space. The harmonic bias potential had a spring constant $\kappa = 2826.5$ kcal/mol. 
The length of the simulations is dependent of the $P, T$ conditions: for 180 K at 2 kbar, 182 K at 1.7 kbar and 160 K at 2 kbar the simulations were 25 ns long. For 170 K at 1 and 3 kbar, the simulations were 50 ns long. For 170 K at 2 kbar the simulation was 60 ns long. For all other simulations, they were 100 ns long. %

Finally, the data accumulated in the different windows was combined together to compute the free energy profile by means of the binless weighted histogram analysis method (called also multistate Bennett acceptance ratio)~\cite{Shirts-jcp-08}, using open source code from Joshua Goings (https://github.com/jjgoings/wham). For comparison, we also reconstructed free-energy profiles using Alan Grossfield's implentation~\cite{Grossfield-wham} of the traditional method in Ref. \cite{Kumar-jcp-92}, finding differences of at most 1 $k_BT$ (see SI Fig. 6).

As a first indication of the convergence of umbrella sampling, we have computed the auto-correlation function of the $S$-path coordinate in umbrella sampling windows:
    $\langle x(0)x(t) \rangle / \langle x^2 \rangle$ with $x(t)=S(t)-\langle S \rangle$,
discarding the first quarter of each trajectory as equilibration. We report the correlation functions in SI Fig. 14, and we also report for comparison the corresponding functions for the density and for the $Q6$ Steinhardt order parameter in SI Fig. 15 and 16, respectively.%
We estimated statistical uncertainties on free-energy profiles using block averages, taking the largest value of the standard error of the mean (see Fig.~\ref{fig:FES}). We cut our trajectory into 2 to 10 blocks, computed the free-energy for each block and estimated the standard error, then we took the largest error among the different numbers of blocks.

\subsection*{Coordination number and LDL/HDL cluster analysis}

We computed the oxygen-oxygen coordination number $C_{OO}$ as the number of neighbors within a cutoff of 0.34 nm, using PLUMED with the following switching function: $\mathsf{c}(r) = (1 - (r/0.34)^{32}) / (1 - (r/0.34)^{64})$. 
Next, we time-averaged the coordination number for each atom over time intervals of 20 ps along the umbrella sampling trajectory and computed the probability distribution (see SI Fig. 13). To reduce irrelevant noise, we smoothed the histogram by averaging over the adjacent bins. To obtain equilibrium populations, we re-weighted the contribution of each umbrella sampling window according to the Boltzmann factor $Z^{-1}\mathrm{e}^{-G(S)/k_BT}$.

In Figure \ref{fig:bulkfraction} we traced the number of bulk molecules in a spherical drop of LDL or HDL water by counting, in the reference structures at 170 K and 0 or 5 kbar, how many molecules belonging to a sphere (mathematically defined within a bulk periodic configuration) are in contact only with molecules of the sphere itself and not with external molecules. Contact is defined using the same switching function $\mathsf{c}(r)$ discussed above.
On the opposite side, as a reference limit for the case of random mixing between LDL and HDL molecules we generated random bond networks with the same distribution of coordination numbers as obtained from MD simulations, starting from a random initial adjacency matrix and adding/removing random bonds ($10^5$ Metropolis Monte Carlo steps) until a deviation  $\int dC_{OO}|P_{MD}(C_{OO})-P_{RN}(C_{OO})|= 0.058\pm 0.007$ between the probability distributions of coordination numbers from MD and from the random network.


\begin{acknowledgments}
  This work was granted access to the Occigen cluster at CINES under allocations No. A004091387 and A005090143 made by GENCI, and to the JeanZay cluster at IDRIS under allocation No. A0060901387.
  We gratefully acknowledge the CPU time provided by PRACE under the project 2018194705.
  We thank Silvio Pipolo for a significant help with the path coordinates implementation in PLUMED and for useful discussions.
  We thank Fr\'ed\'eric Caupin, Stefan Klotz, Fr\'ed\'eric Datchi, Livia Bove, Fr\'ed\'eric van Wijland, and Carlos Vega for useful discussions.
\end{acknowledgments}



\bibliography{biblio}

\newpage
\clearpage

\section*{Supporting Information}

\setcounter{table}{0}
\renewcommand{\thetable}{S\arabic{table}}%
\setcounter{figure}{0}
\renewcommand{\thefigure}{S\arabic{figure}}%


\begin{figure*}[!ht]
    \centering
    \includegraphics[width=0.6\linewidth]{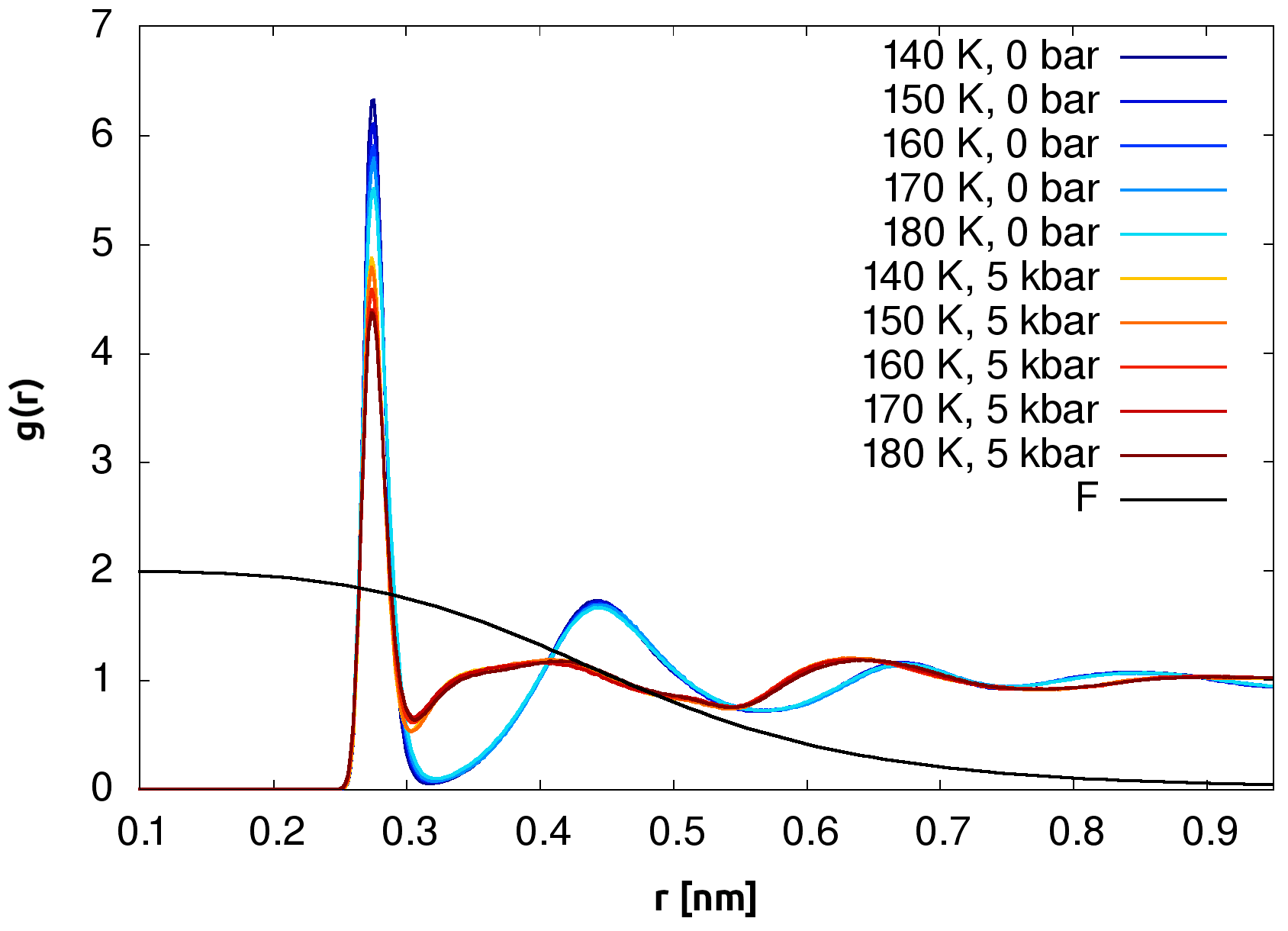}
    \caption{Radial distribution functions $g(r)$ of the low-density ($P=0$ bar) and high-density ($P=5$ kbar) reference states at different temperatures and switching function $F(r)$ (multiplied by 2 for easier visualization). The reference states and $F(r)$ are employed to define the PIV metric (see Materials and Methods for details).}
    \label{fig:rdf_switch}
\end{figure*}

\begin{figure*}[!ht]
    \centering
    \includegraphics[width=0.6\linewidth]{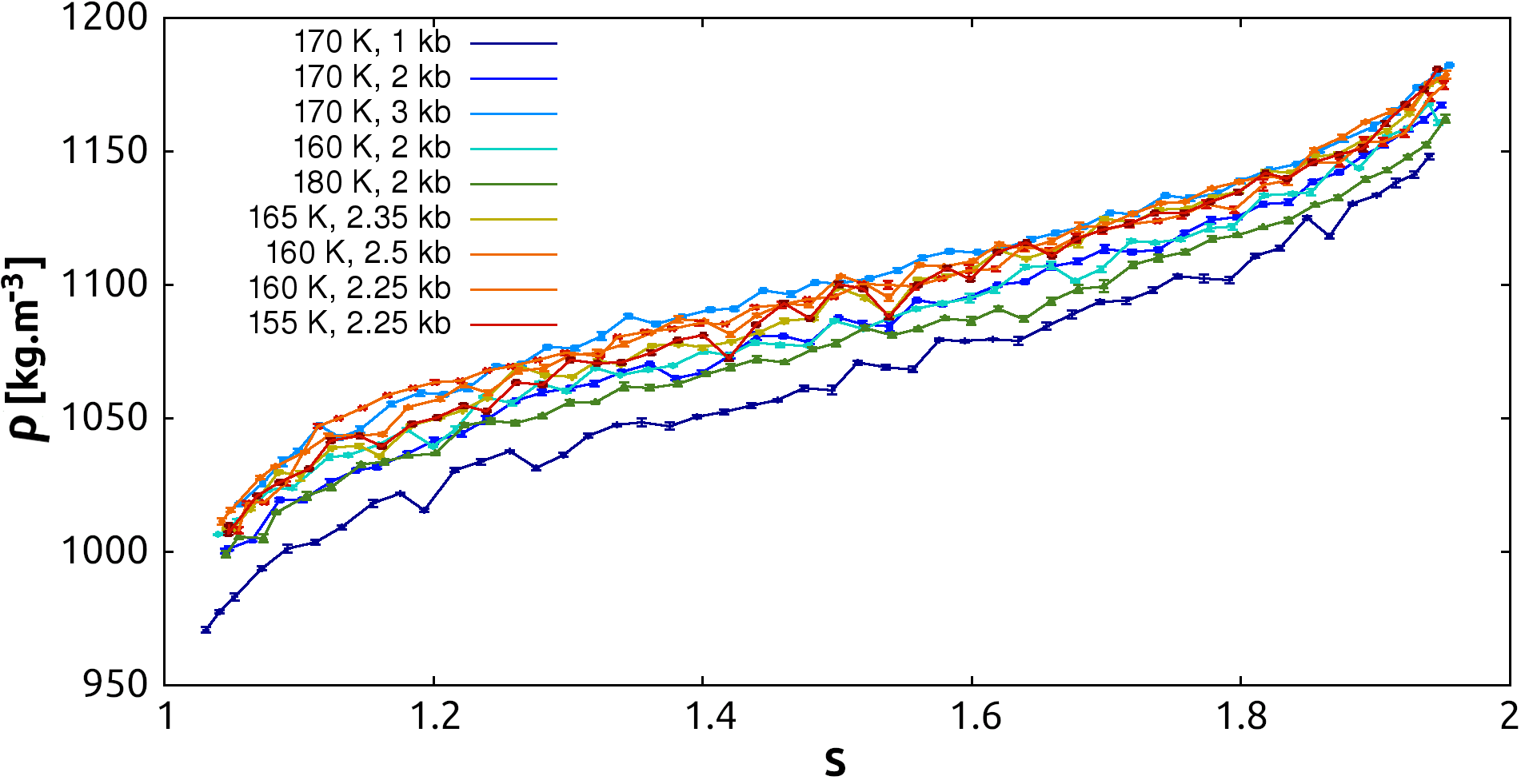}
    \caption{Average density 
    as a function of the $S$ path coordinate for every umbrella sampling window for various $P,T$ conditions. An almost linear correlation can be observed. The bars indicate the standard deviation of the density.}
    \label{fig:correlation_density_S}
\end{figure*}



\begin{table*}
\centering
\caption{Average box vectors starting from low-density state (from unbiased shooting trajectories, second half of trajectories)}
\label{table:low_dens_box}
  \begin{tabular}{ccccccc}
    $P, T$ &
    $ \langle A \rangle $ ({\AA}) &
    $ \langle B \rangle $ ({\AA}) &
    $ \langle C \rangle $ ({\AA}) &
    $ \langle\alpha\rangle $ & $ \langle\beta\rangle $ & $ \langle\gamma\rangle $ \\
    \midrule
    160 K, 2.5 kbar &
        $24.7 \pm 0.1$ & $23.2 \pm 0.1$ & $42.3 \pm 0.2$ &
        $100$ & $79$ & $110$\\
    170 K, 2 kbar &
        $24.6 \pm 0.1$ & $23.2 \pm 0.1$ & $43.0 \pm 0.2$ &
        $100$ & $79$ & $109$ \\
    180 K, 2 kbar &
        $22.8 \pm 0.1$ & $28.1 \pm 0.1$ & $38.0 \pm 0.1$ &
        $96$ & $83$ & $101$ \\
    \bottomrule
  \end{tabular}
\centering
\caption{Average box vectors starting from high-density state (from unbiased shooting trajectories, second half of trajectories)}
\label{table:high_dens_box}
  \begin{tabular}{ccccccc}
    $P, T$ &
    $ \langle A \rangle $ ({\AA}) &
    $ \langle B \rangle $ ({\AA}) &
    $ \langle C \rangle $ ({\AA}) &
    $ \langle\alpha\rangle $ & $ \langle\beta\rangle $ & $ \langle\gamma\rangle $ \\
    \midrule
    160 K, 2.5 kbar &
        $24.0 \pm 0.1$ & $23.6 \pm 0.1$ & $38.9 \pm 0.1$ &
        $96$ & $87$ & $106$\\
    170 K, 2 kbar &
        $24.7 \pm 0.1$ & $23.8 \pm 0.1$ & $39.5 \pm 0.1$ &
        $95$ & $83$ & $107$ \\
    180 K, 2 kbar &
        $22.3 \pm 0.1$ & $28.6 \pm 0.1$ & $36.4 \pm 0.1$ &
        $89$ & $95$ & $109$ \\
    \bottomrule
  \end{tabular}
  
\end{table*}


\begin{figure*}[!ht]
    \centering
    \includegraphics[width=0.7\linewidth]{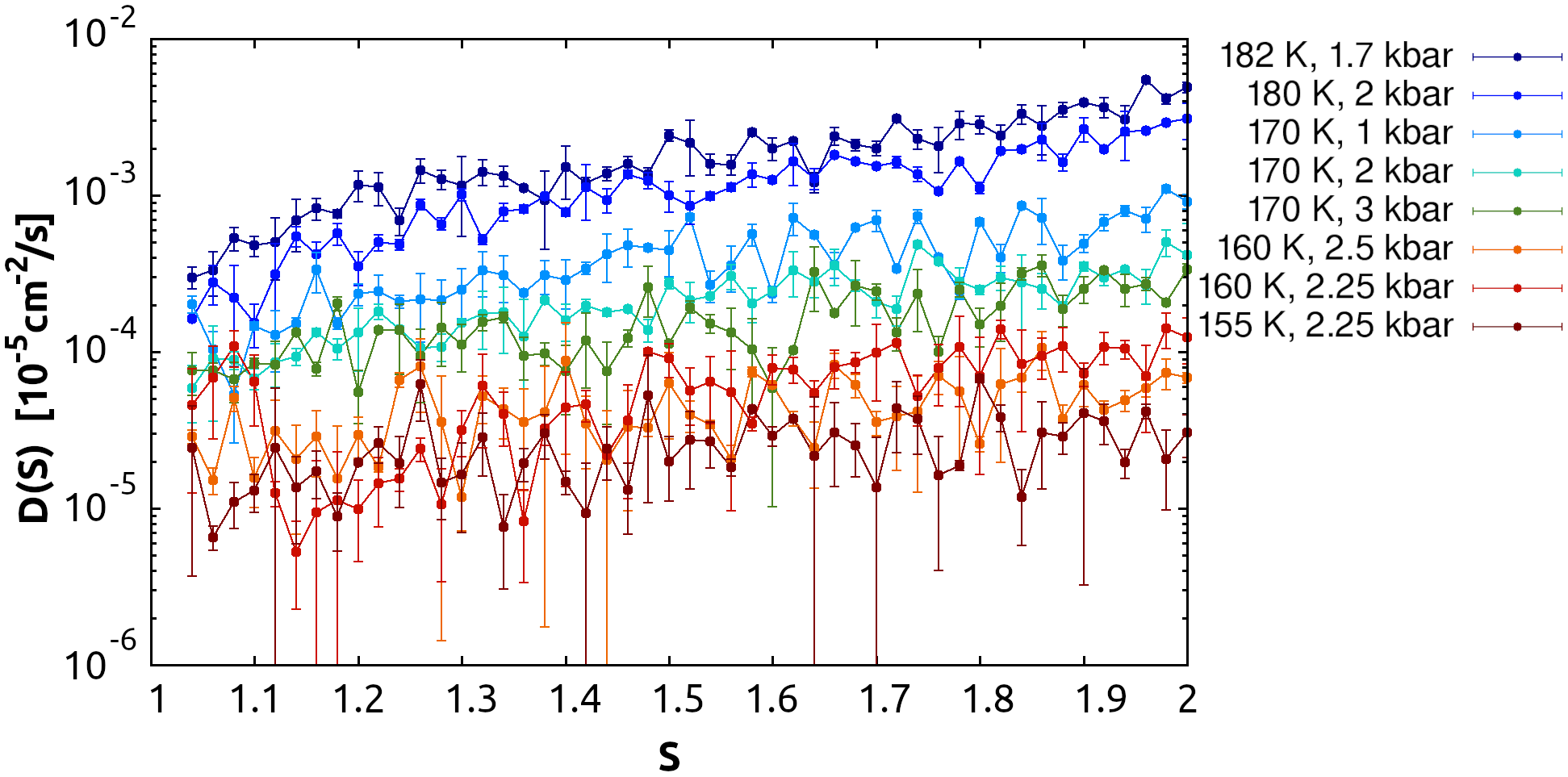}
    \caption{
    Diffusion coefficients of oxygen atoms for several $P,T$ conditions, computed from the mean square displacement with gromacs on the biased umbrella sampling trajectories for each windows. The first half of the trajectory is discarded as equilibration. The error bar represent the standard deviation.
    }
    \label{fig:diffusion_coeff}
\end{figure*}

\begin{figure*}[!ht]
    \centering
    \includegraphics[width=0.95\linewidth]{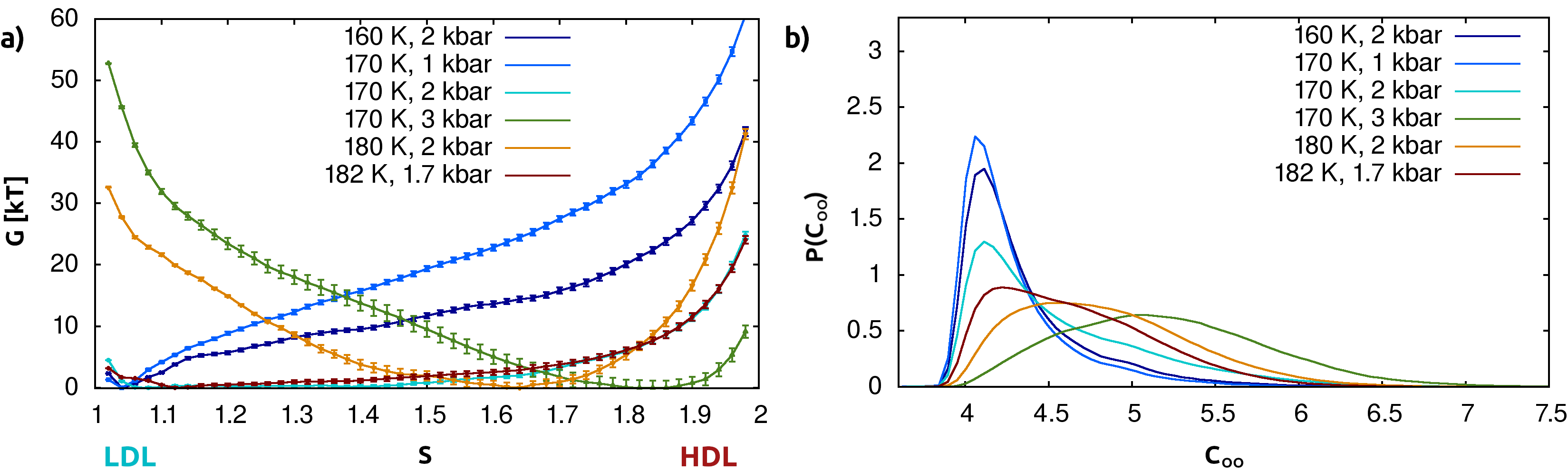}
    \caption{(a) Free-energy profiles for the low-density/high-density water transformation, and (b) corresponding distributions of oxygen-oxygen coordination numbers $C_{OO}$, computed using an isotropic Berendsen barostat with a relaxation time of 0.5 ps. Umbrella sampling have a length of 15 ns for all points except 182 K, 1.7 kbar where it is of 25 ns. The distribution of $C_{OO}$ is computed in the same way as for the Parinello-Rahman barostat, see its definition in Material and Methods. $S\approx1.1$ correspond to low density and $S\approx1.9$ to high density.}
    \label{fig:comp_pari_beren}
\end{figure*}

\begin{figure*}[!ht]
    \centering
    \includegraphics[width=0.5\linewidth]{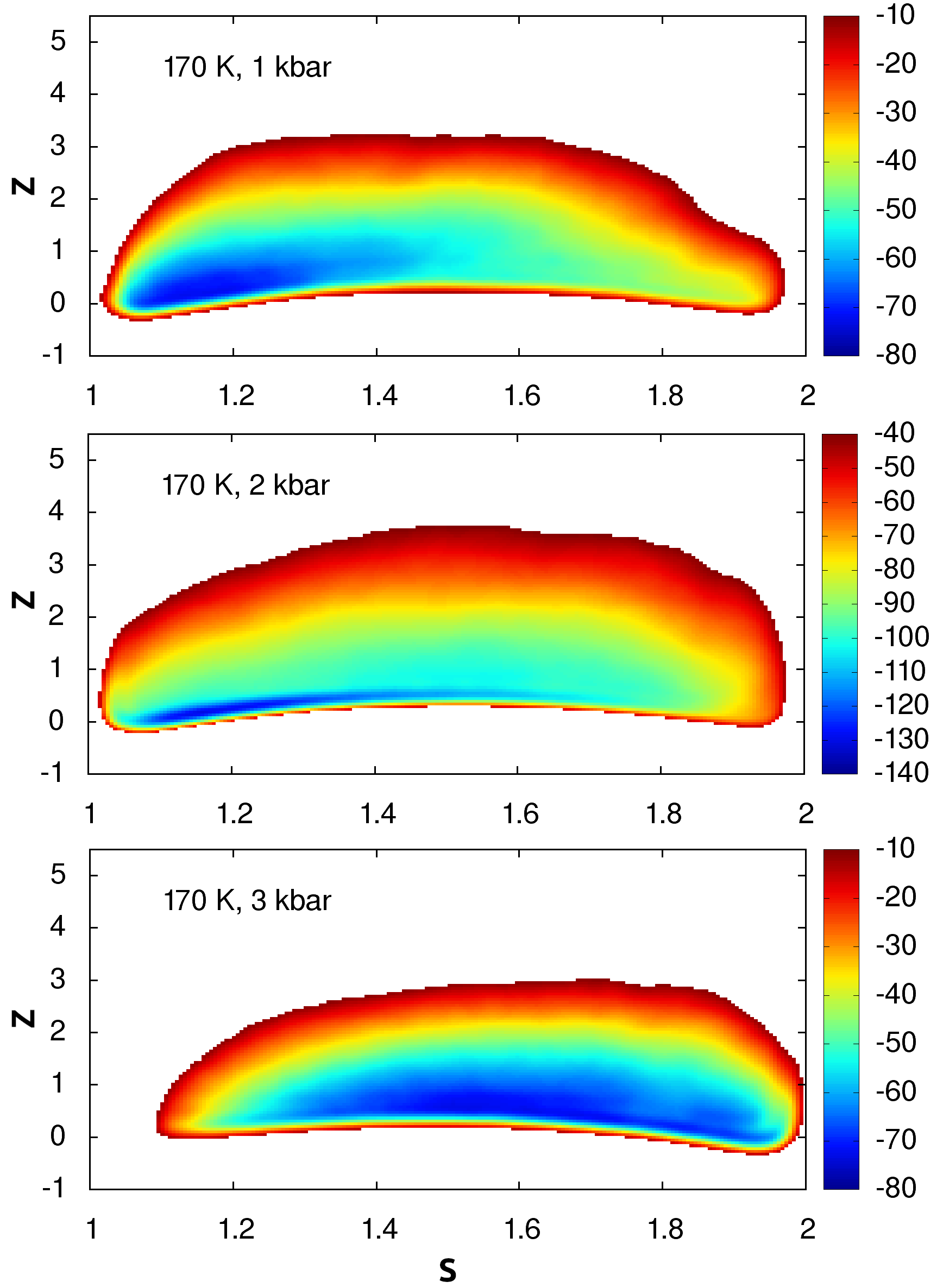}
    \caption{Bias surface reconstructed from metadynamics at 170 K and 1, 2 and 3 kbar as a function of the  PIV-based path coordinates $S$ and $Z$ (see Materials and Methods for details). The color scale is in $k_BT$ units. No significant feature can be observed along the $Z$ direction.}
    \label{fig:F-meta}
\end{figure*}

\begin{figure*}[!ht]
    \centering
    \includegraphics[width=\linewidth]{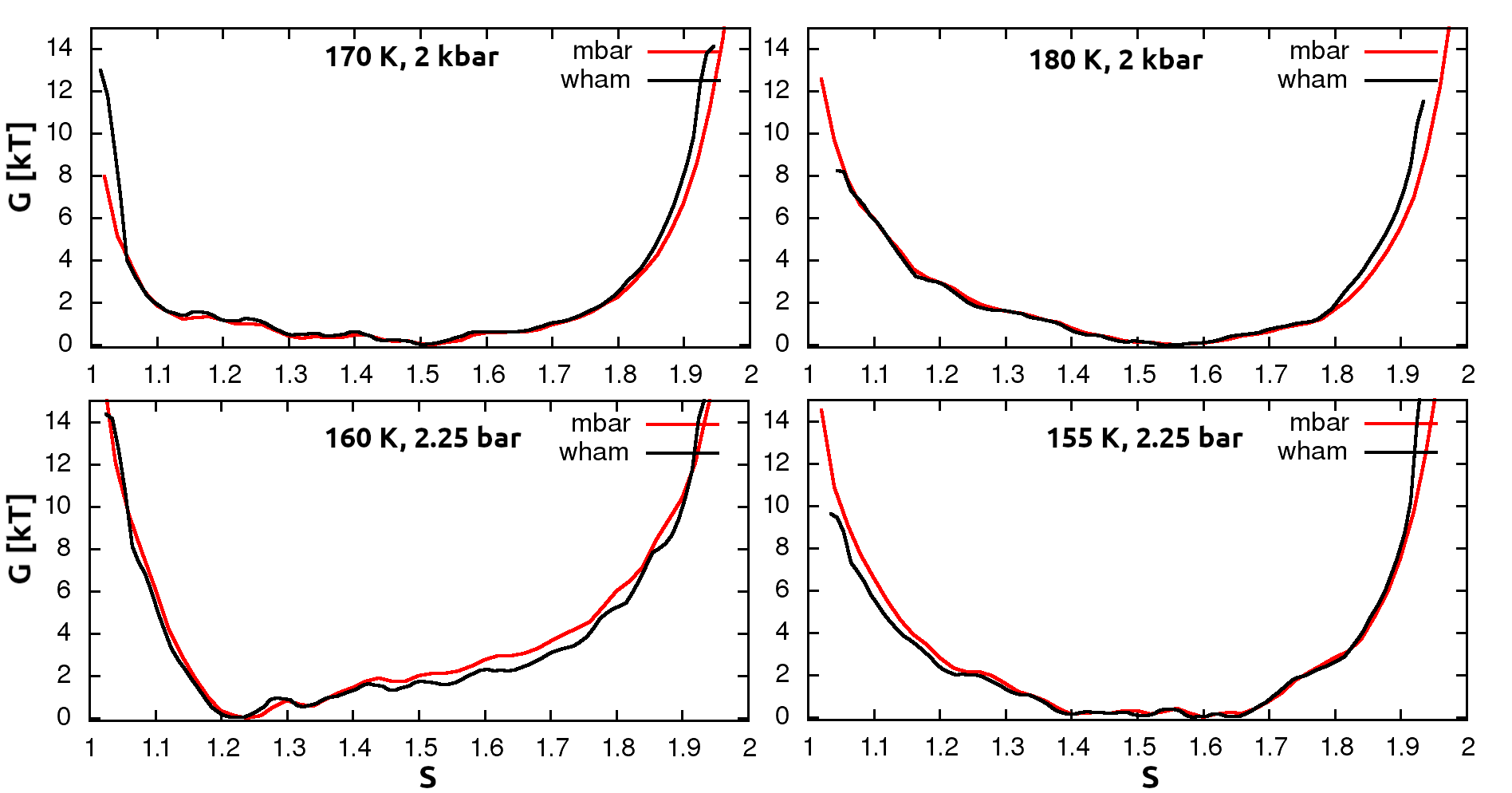}
    \caption{Comparison of two techniques used to compute free energy profiles from umbrella sampling simulations: in red the Multistate Bennett Acceptance Ratio (MBAR) - binless Weighted Histogram Analysis Method (binless-WHAM) \cite{Shirts-jcp-08,Goings-wham}, and in black the WHAM from Ref. \cite{Kumar-jcp-92,Grossfield-wham}. The first quarter of each trajectory was discarded as equilibration.}
    \label{fig:mbar_wham_comp}
\end{figure*}



\begin{table*}
  \centering
  \begin{tabular}{crr}
    $P, T$ &
    $D$ ($10^{-5}$ cm$^2$/s), $S$ = 1.02 &
    $D$ ($10^{-5}$ cm$^2$/s), $S$ = 1.98 \\
    \midrule
    155 K, 2.25 kbar &
      $8 \pm 2\; 10^{-6}$ &
      $9 \pm 1\; 10^{-6}$ \\
    160 K, 2.5 kbar &
      $2.0 \pm 0.7\; 10^{-5}$ &
      $3.1 \pm 0.8\; 10^{-5}$ \\
    170 K, 2 kbar &
      $1.0 \pm 0.4\; 10^{-4}$ &
      $1.8 \pm 0.6\; 10^{-4}$ \\
    180 K, 2 kbar &
      $1.2 \pm 0.1\; 10^{-3}$ &
      $1.12 \pm 0.07\; 10^{-3}$ \\
    \bottomrule
  \end{tabular}
  \caption{Diffusion coefficients of oxygen atoms for several $P,T$ conditions, computed from the mean square displacement with gromacs on unbiased MD trajectories of 500 ns or more. The first half of the trajectory is discarded as equilibration. The two columns present average values from trajectories starting from low- ($S=1.02$) or high-density ($S=1.98$) states.}
  \label{table:diffusion}
  
\end{table*}

\begin{figure*}[!ht]
    \centering
    \includegraphics[width=0.6\linewidth]{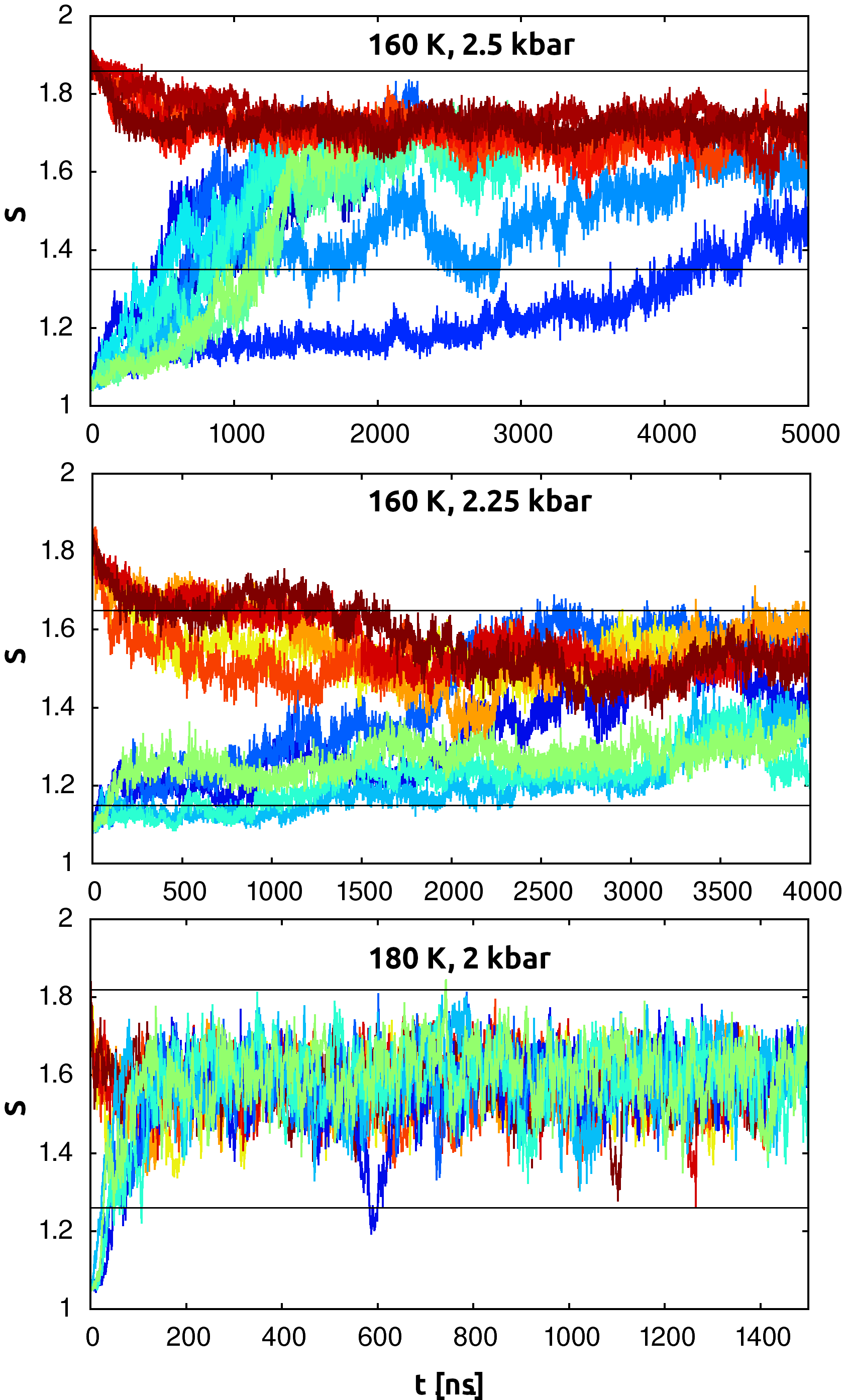}
    \caption{
    Independent unbiased MD trajectories initiated from LDL (blue) or HDL (red) end-point umbrella sampling configurations. 
    The intervals delimited by black lines correspond to free-energy values within 2 $k_BT$ from the free-energy minimum as reconstructed from umbrella sampling (Figure~1.a-c in the main text). 
    Both the initial high-density and low-density configurations relax within the very broad free-energy basin of intermediate density, displaying slow diffusion. 
    At 160 K and 2.25-2.5 kbar, the relaxation occurs on the $\mu$s timescale, while at 180  K and 2 kbar it is one order of magnitude faster.
    }
    \label{fig:unbiased_sup}
\end{figure*}


\begin{figure*}[!ht]
    \centering
    \includegraphics[width=0.7\linewidth]{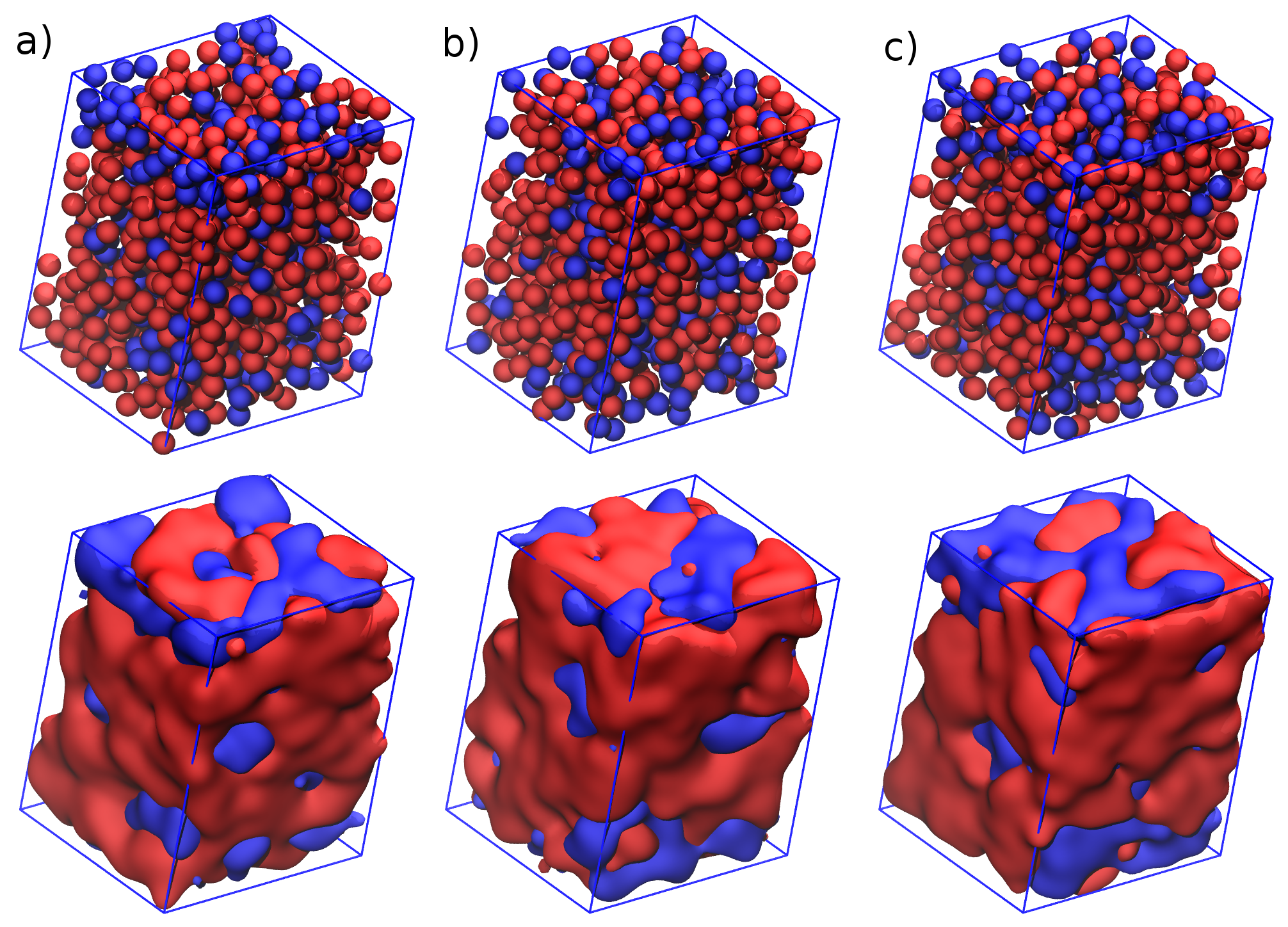}
    \caption{
        Example of 3D structures from unbiased MD at 160 K, 2.25 kbar. LDL-like molecules (with $C_{OO}<4.5$) are indicated as blue balls, HDL-like ones (with $C_{OO}>4.5$) as red balls; surfaces are drawn with the "QuickSurf" tool of vmd (isosurfaces extracted from a volumetric Gaussian density maps on a uniformly-spaced 3-D lattice) \cite{vmd-jmg-96-article,vmd-quicksurf}.
    }
    \label{fig:3D}
\end{figure*}

\begin{figure*}[!ht]
    \centering
    \includegraphics[width=0.9\linewidth]{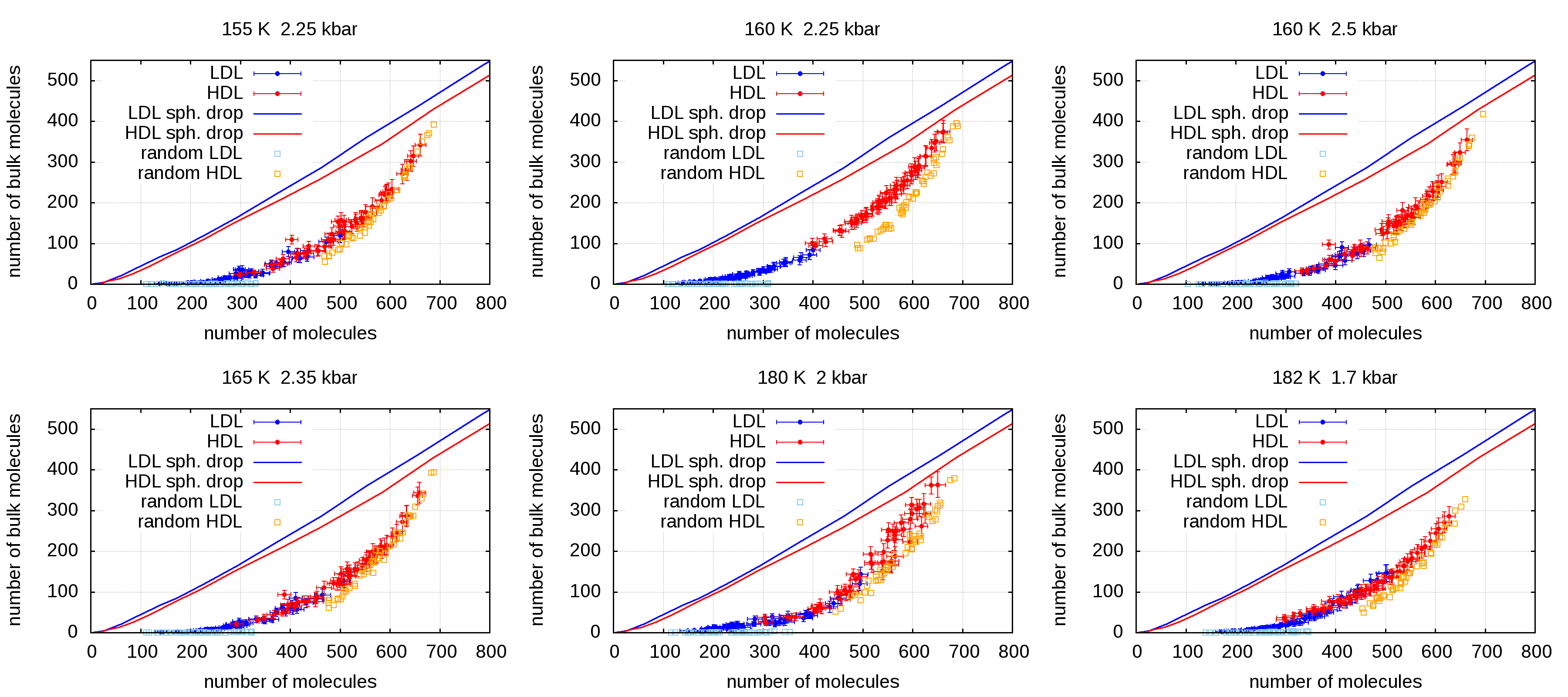}
    \caption{
    Number of LDL-like or HDL-like molecules (assigned on the basis of $C_{OO}<4.5$ or $>4.5$, respectively) that are surrounded by molecules of the same type, i.e. not at the interface LDL/HDL, within umbrella sampling trajectories. For comparison, continuous curves indicates the number of bulk molecules in a spherical droplet containing only alike molecules, and the squares correspond to random networks of LDL-like or HDL-like molecules with the same bond distribution as MD configurations (see Materials and Methods for details).
    }
    \label{fig:USbulkfraction}
\end{figure*}


\begin{figure*}[!ht]
    \centering
    \includegraphics[width=0.7\linewidth]{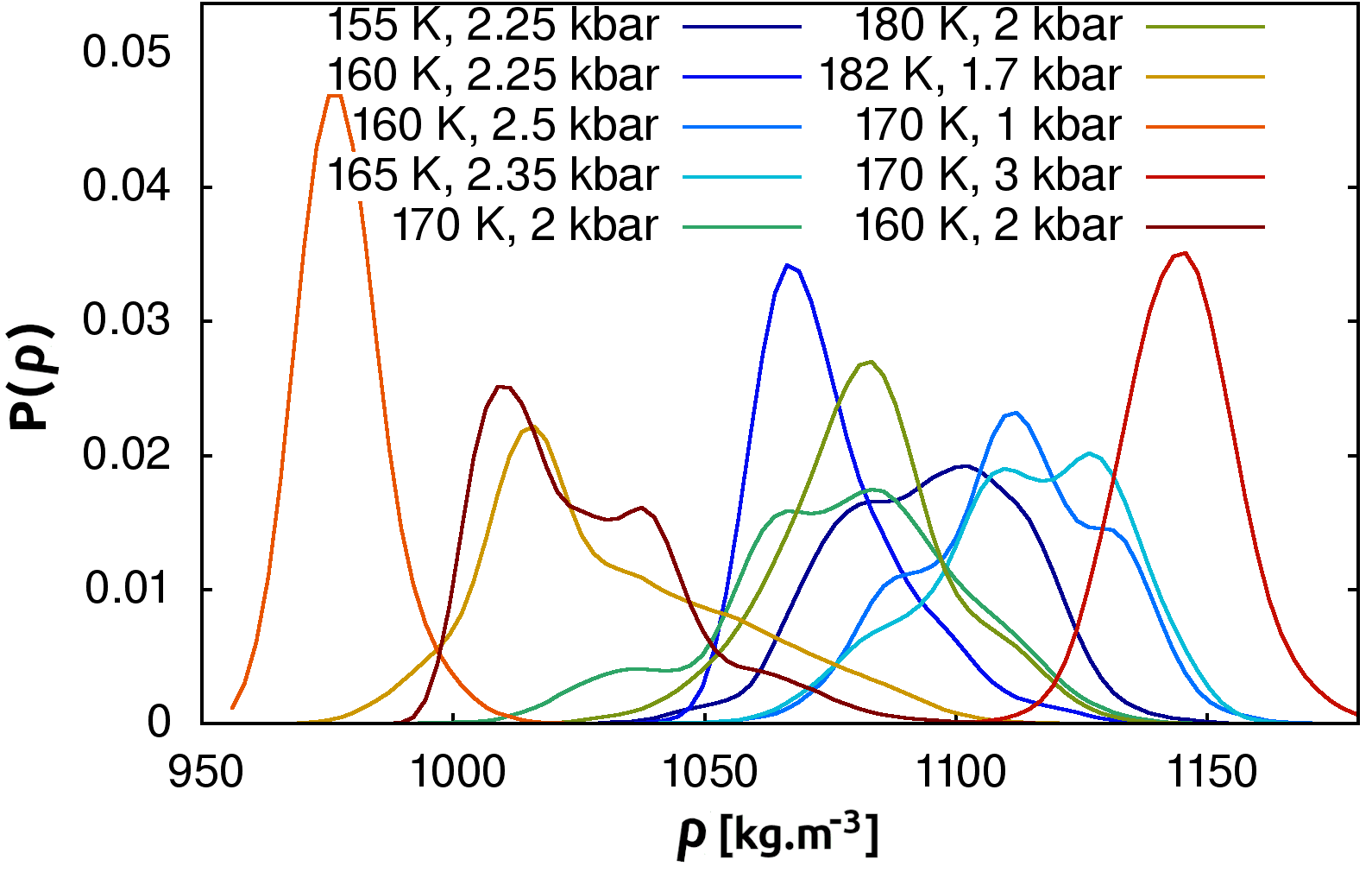}
    \caption{Distribution of density $\rho$ for the various $(P,T)$ condition explored in this study, computed from the weighted contribution of every umbrella sampling window with the Boltzmann factor $\mathrm{e}^{-G(S)/k_BT}$ to obtain equilibrium distributions. To reduce irrelevant noise, we smoothed the histogram by averaging over the adjacent bins.}
    \label{fig:PT_density_distri}
\end{figure*}

\begin{figure*}[!ht]
    \centering
    \includegraphics[width=0.7\linewidth]{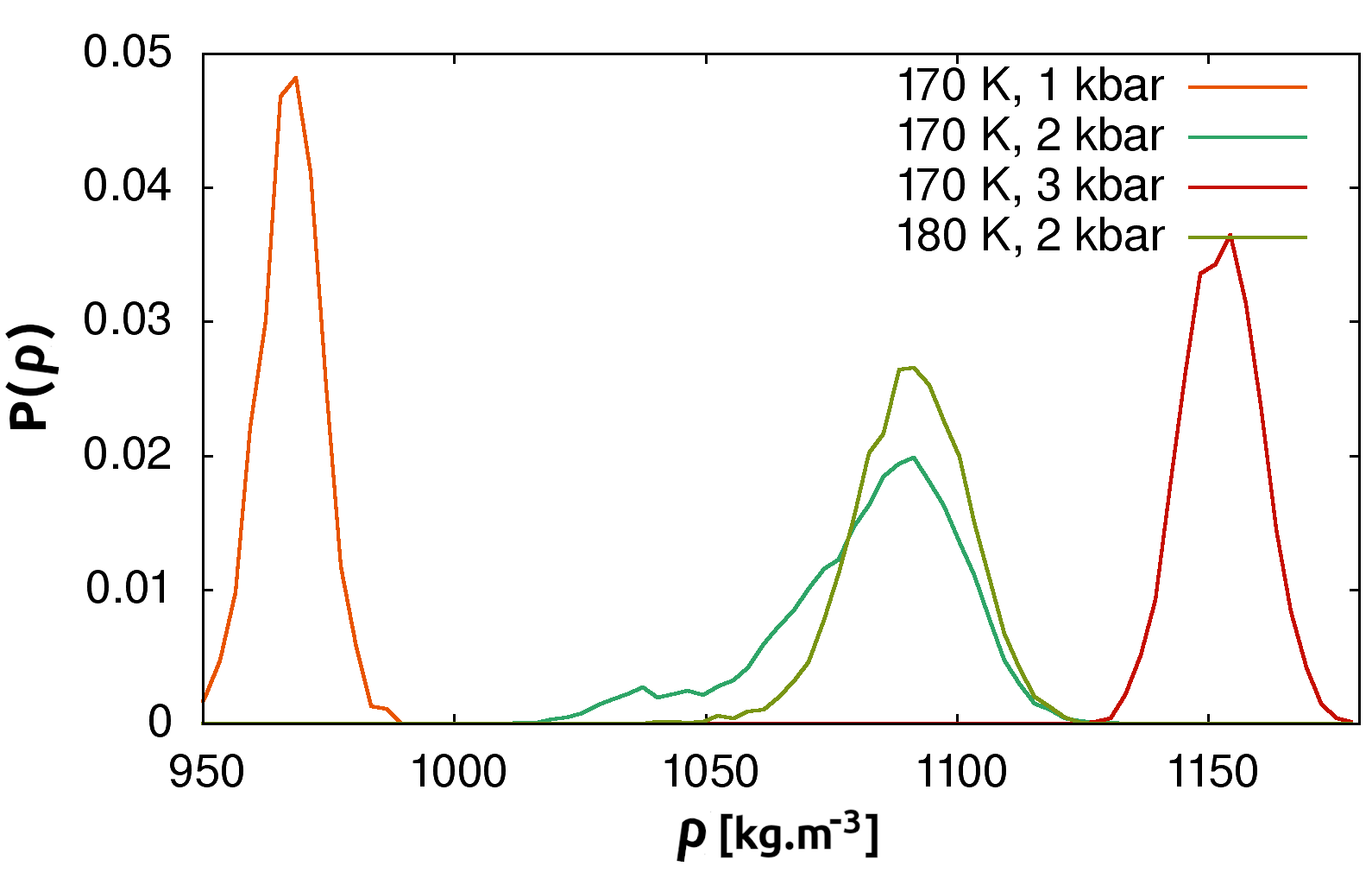}
    \caption{Distribution of density $\rho$ for some $(P,T)$ condition explored in this study, estimated from unbiased simulations presented in main text Fig. 3 and SI Fig. 7. The first half of the trajectories is discarded as equilibration. 
    The distributions are consistent with those computed using umbrella sampling data in SI Fig. 10}
    \label{fig:PT_density_distri_shootings}
\end{figure*}

\begin{figure*}[!ht]
    \centering
    \includegraphics[width=\linewidth]{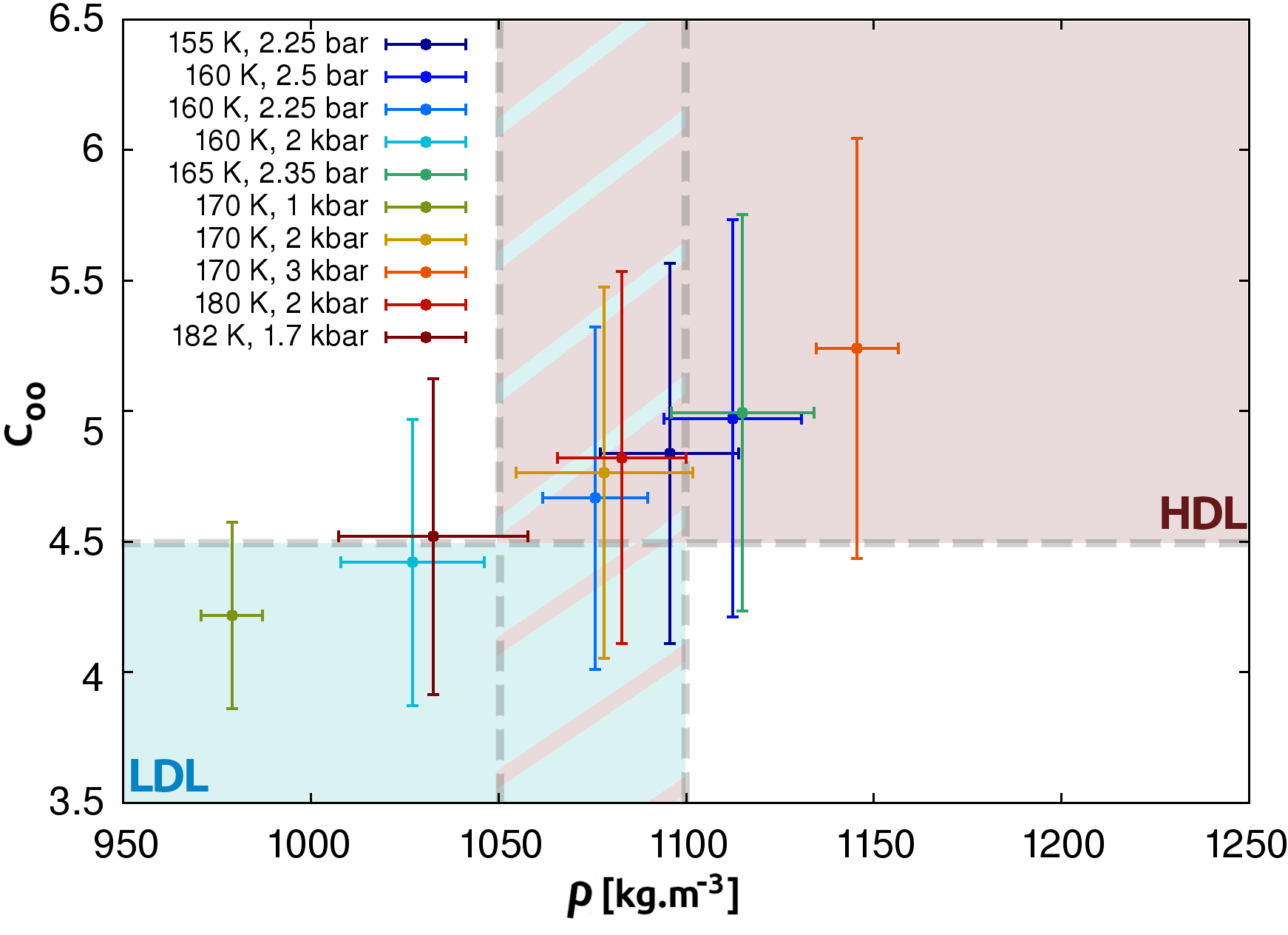}
    \caption{Average oxygen-oxygen coordination number $C_{\mathrm{OO}}$ as a function of the average density $\rho$ for the various $(P,T)$ condition explored in this study, computed from the weighted contribution of every umbrella sampling window with the Boltzmann factor $\mathrm{e}^{-G(S)/k_BT}$ to obtain equilibrium distribution. The respective fluctuations are shown as horizontal and vertical bars (standard deviation of the distributions). The horizontal dashed line indicate the criterion used to separate low and high density liquid, according to coordination number~\cite{Harrington-prl-97}. The two vertical dashed lines indicate the extreme values of the density at $S = 1.5$ as shown in SI Fig. 2, as a criterion to separate LDL and HDL regions.}
    \label{fig:PT_CN_rho}
\end{figure*}

\begin{figure*}[!ht]
    \centering
    \includegraphics[width=0.6\linewidth]{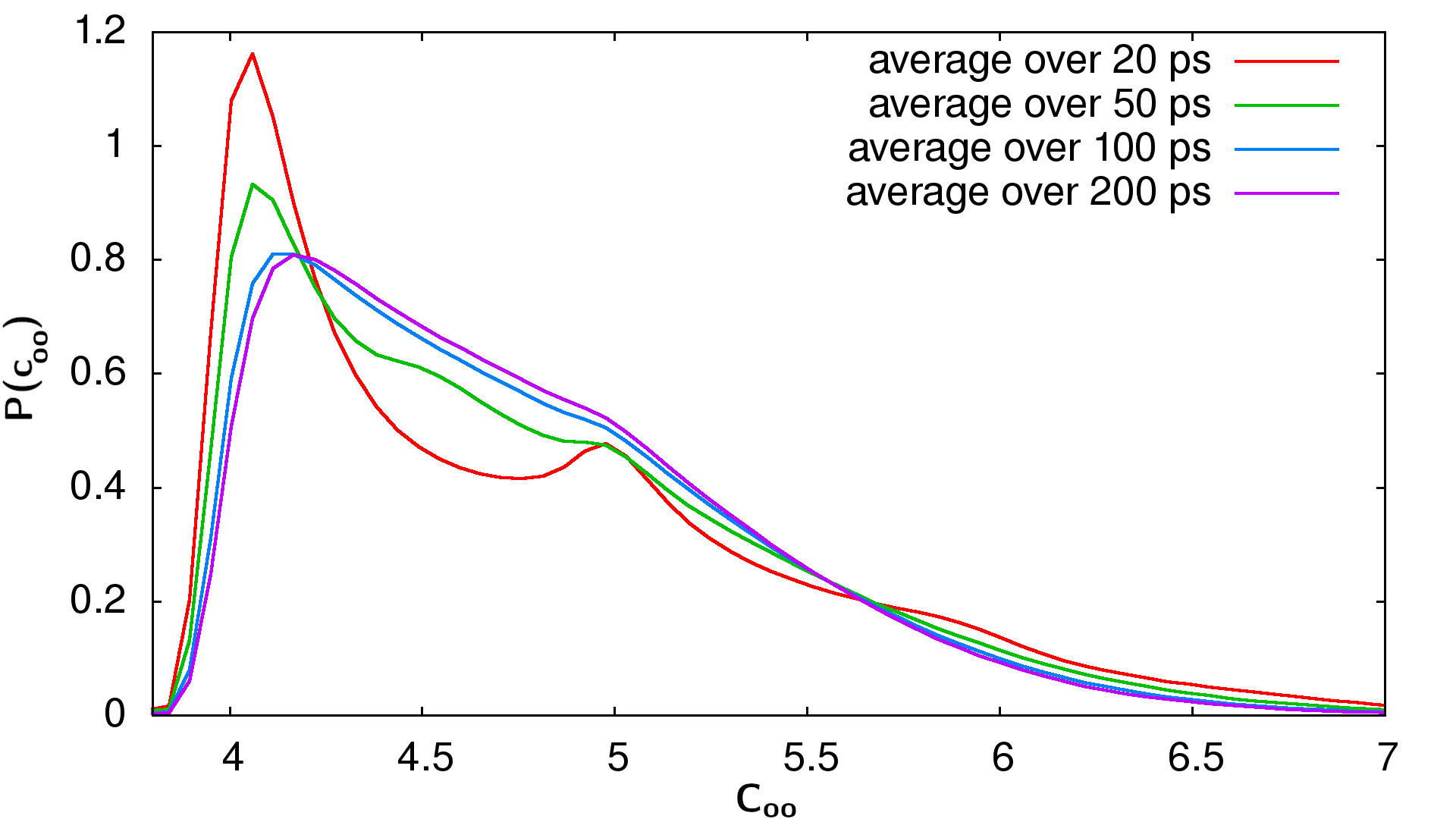}
    \caption{Coordination number probability at 170 K and 2 kbar from umbrella sampling simulations averaging $C_{OO}$ of each oxygen atom over four different time intervals.}
    \label{fig:CN_avg_time}
\end{figure*}


\begin{figure*}[!ht]
    \centering
    \includegraphics[width=0.95\linewidth]{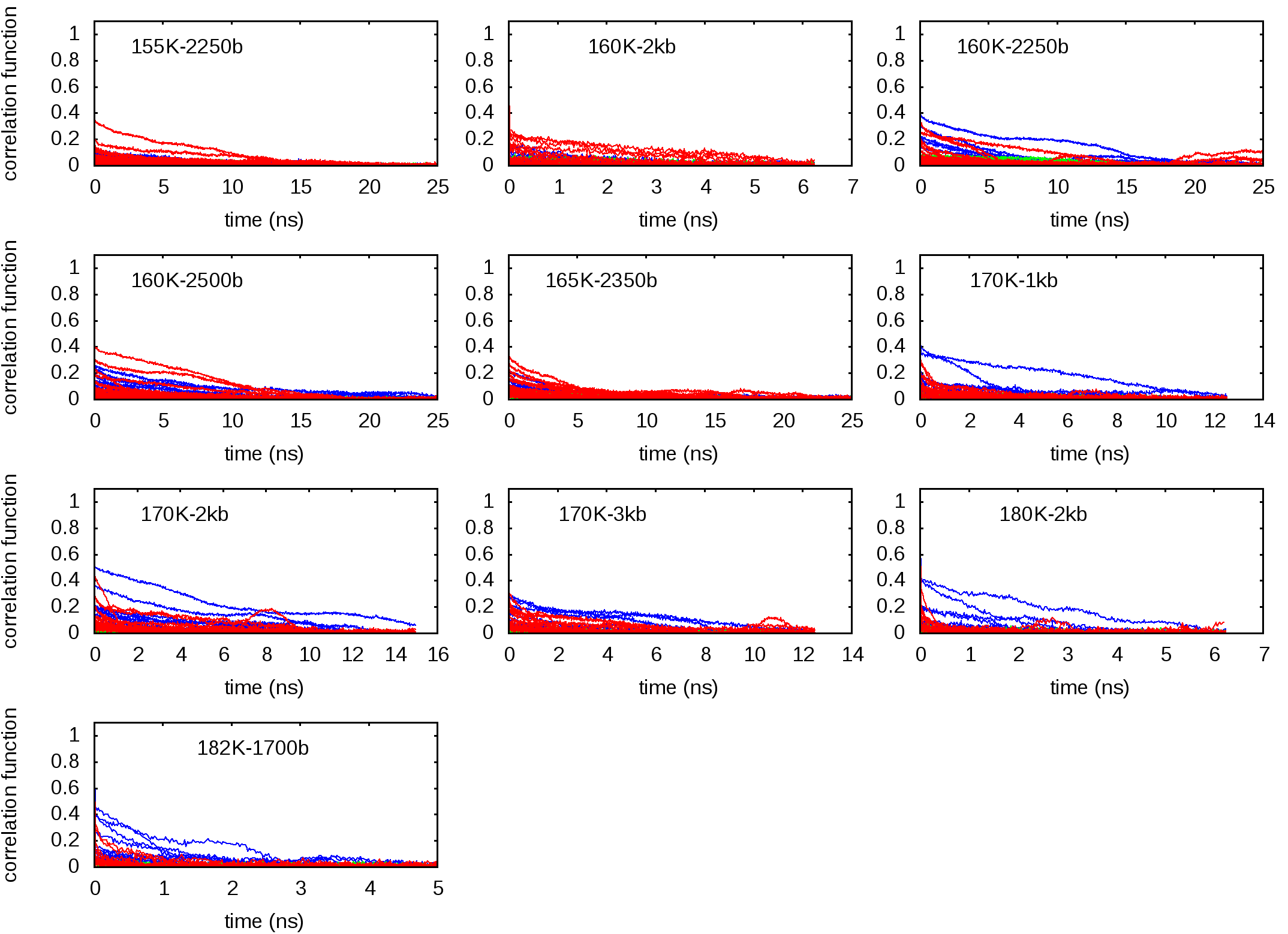}
    \caption{Normalized self-correlation function $\langle x(t)x(0) \rangle / \langle x^2 \rangle$, $x(t)=S(t)-\langle S \rangle$ of the PIV-based path coordinate in umbrella sampling simulations. Each line corresponds to an umbrella sampling window, with $S<1.4$ in blue, $1.4<S<1.6$ in green, and $S>1.6$ in red. The first fourth of each trajectory is not employed to compute the correlation, and the total length of each trajectory is four times the duration plotted.}
    \label{fig:corr_func_s}
\end{figure*}

\begin{figure*}[!ht]
    \centering
    \includegraphics[width=0.95\linewidth]{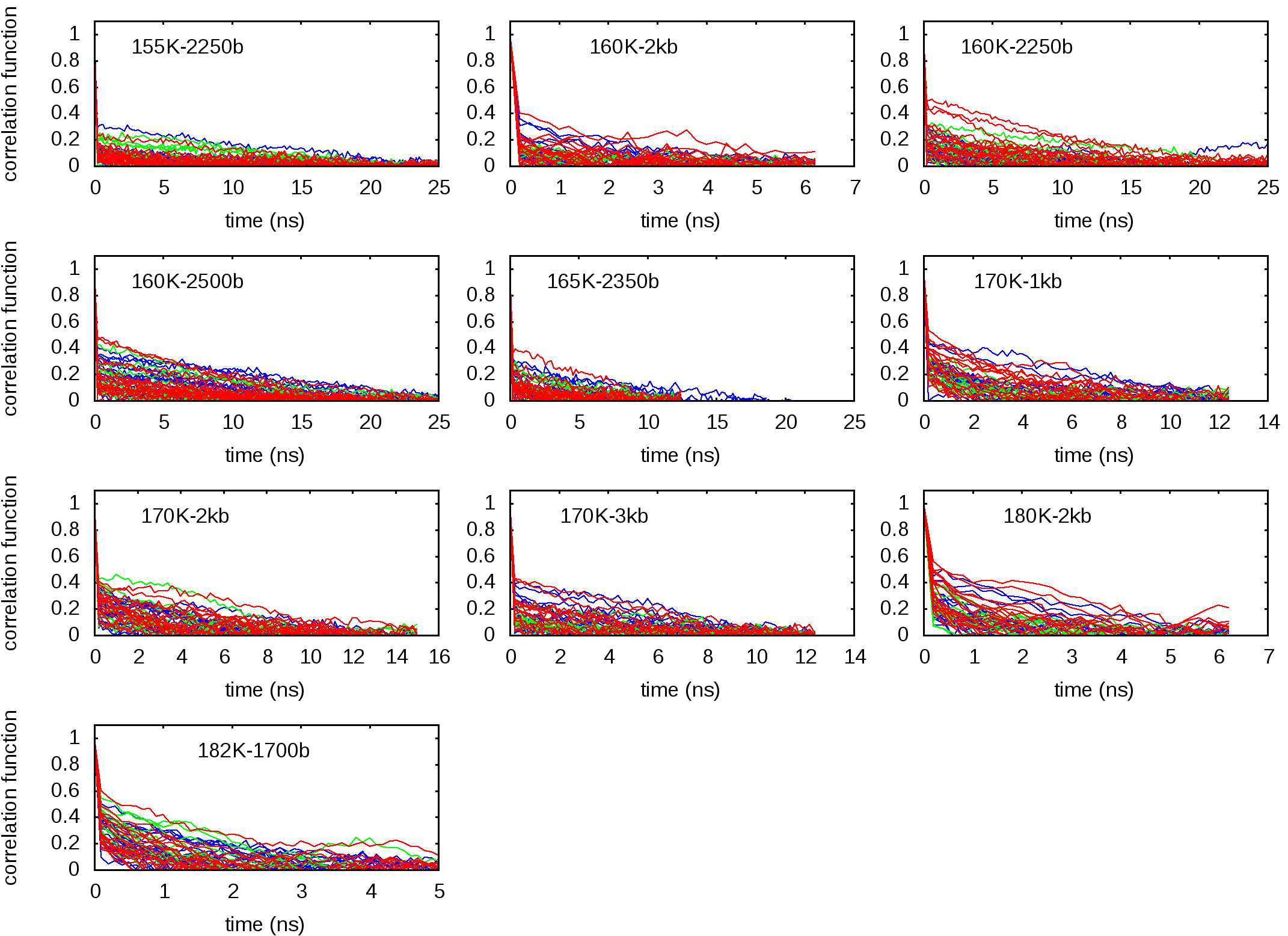}
    \caption{Normalized self-correlation function $\langle x(t)x(0) \rangle / \langle x^2 \rangle$, $x(t)=\rho(t)-\langle \rho \rangle$ of the density in umbrella sampling simulations. Each line corresponds to an umbrella sampling window, with $S<1.4$ in blue, $1.4<S<1.6$ in green, and $S>1.6$ in red. The first fourth of each trajectory is not employed to compute the correlation, and the total length of each trajectory is four times the duration plotted.}
    \label{fig:corr_func_rho}
\end{figure*}

\begin{figure*}[!ht]
    \centering
    \includegraphics[width=0.95\linewidth]{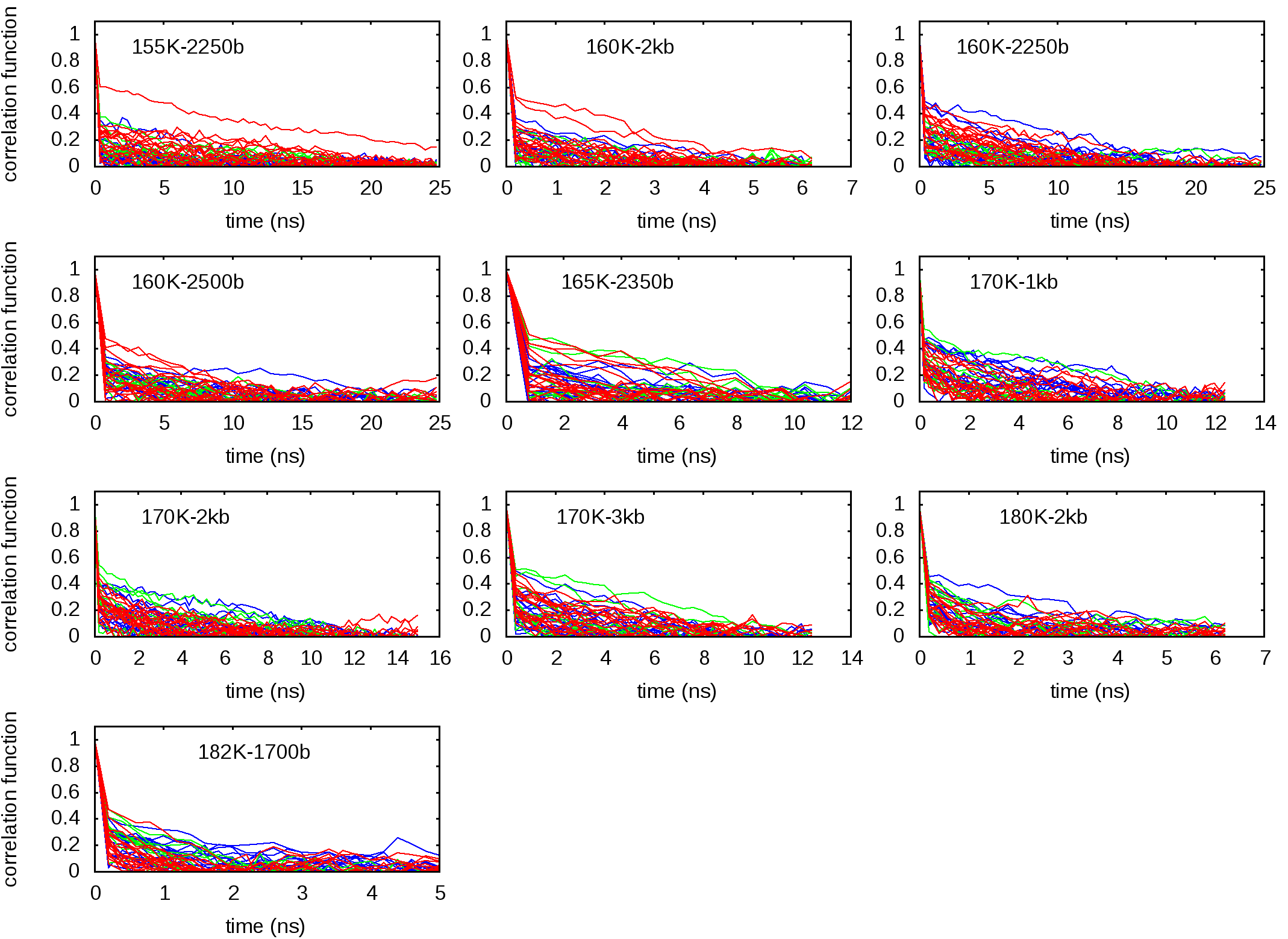}
    \caption{Normalized self-correlation function $\langle x(t)x(0) \rangle / \langle x^2 \rangle$, $x(t)=Q_6(t)-\langle Q_6 \rangle$ of the $Q_6$ Steinhardt order parameter computed for a shell radius of 0.35 nm and averaged over all water molecules in umbrella sampling simulations~\cite{steinhardt-prb-83-parameters}. Each line corresponds to an umbrella sampling window, with $S<1.4$ in blue, $1.4<S<1.6$ in green, and $S>1.6$ in red. The first fourth of each trajectory is not employed to compute the correlation, and the total length of each trajectory is four times the duration plotted.}
    \label{fig:corr_func_q6}
\end{figure*}

\begin{figure*}[!ht]
    \centering
    \includegraphics[width=0.9\linewidth]{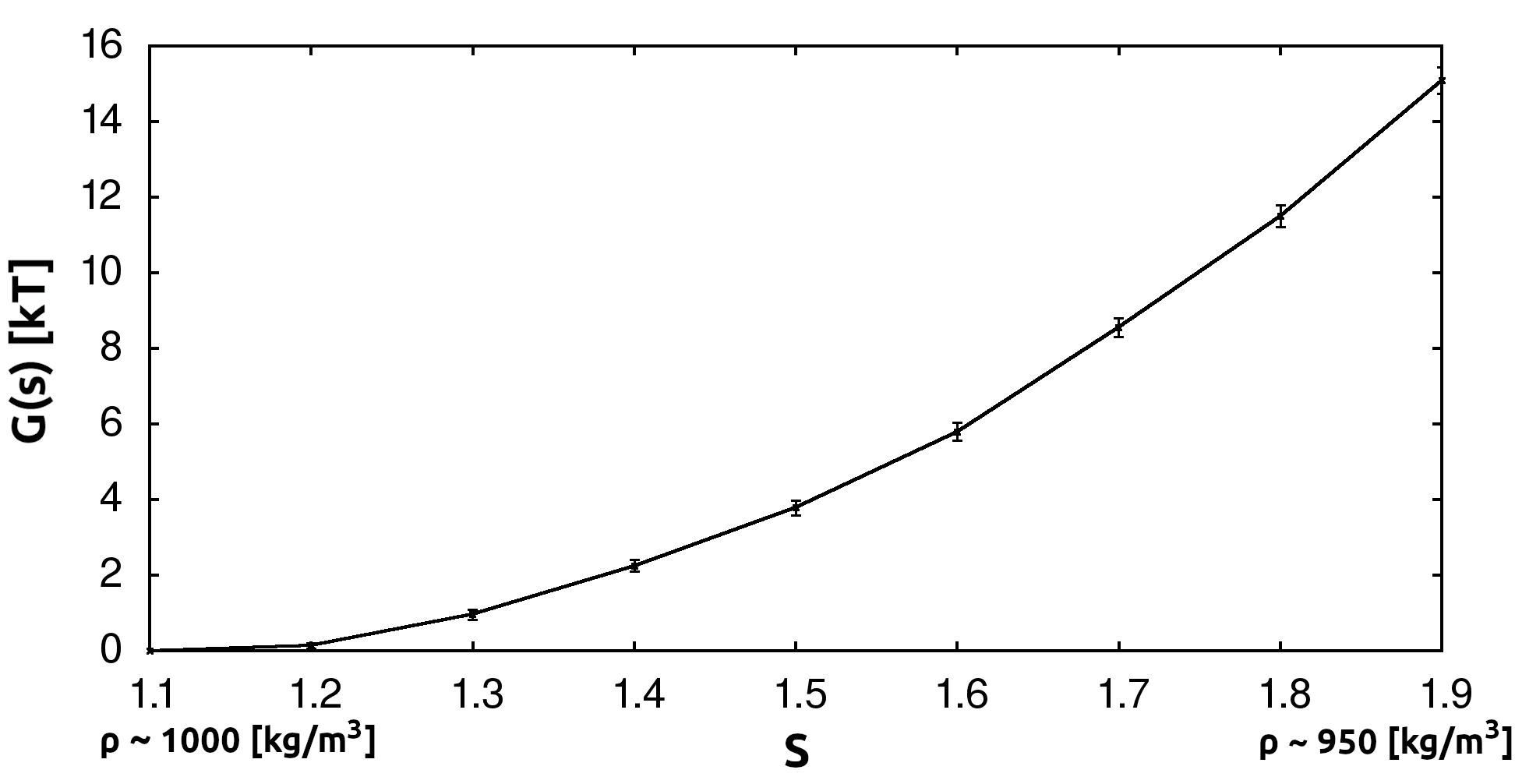}
    \caption{
      Free energy profile at 170 K, 2 kbar between two low-density reference states: the first at $\rho=1000\; \mathrm{ kg.m}^{-3}$, extracted from the umbrella sampling simulation at 170K, 2 kbar reported in the main text, the second at $\rho=950\; \mathrm{ kg.m}^{-3}$, equilibrated at 0 bar. 
      We followed the same protocol as for the other free energy profiles, with 9 umbrella sampling windows, each of a duration of 50 ns and with a spring constant $\kappa = 283.3$ kcal/mol. 
      The error bars are estimated with block averages as explained in Materials and Methods.
    }
\end{figure*}

\begin{figure*}[!ht]
  \centering
  \begin{subfigure}{0.9\linewidth}
    \centering
    \includegraphics[width=0.9\linewidth]{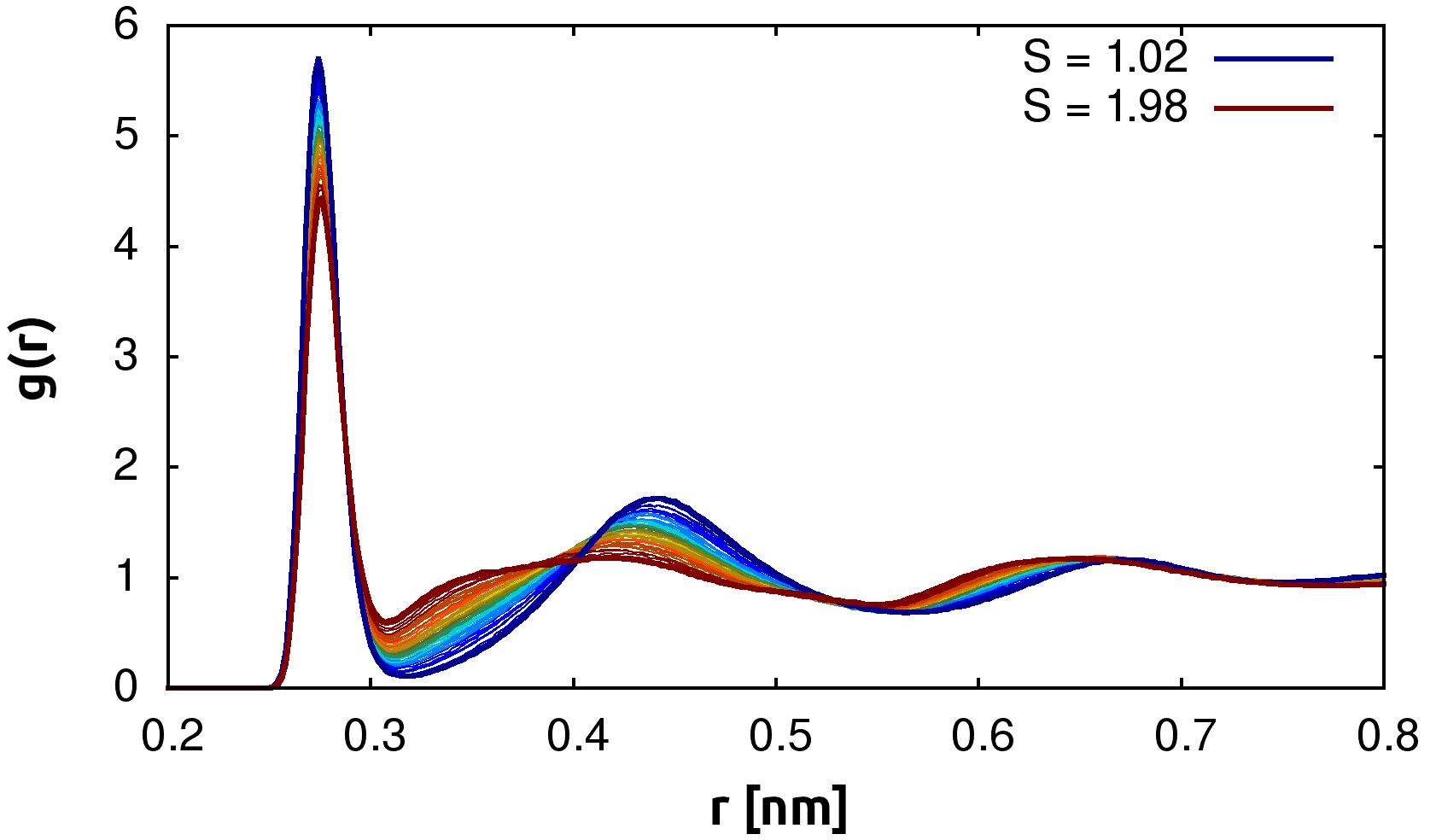}
    \caption{}
    \label{fig:rdf_170K-2kb}
  \end{subfigure}
  \begin{subfigure}{0.9\linewidth}
    \centering
    \includegraphics[width=0.9\linewidth]{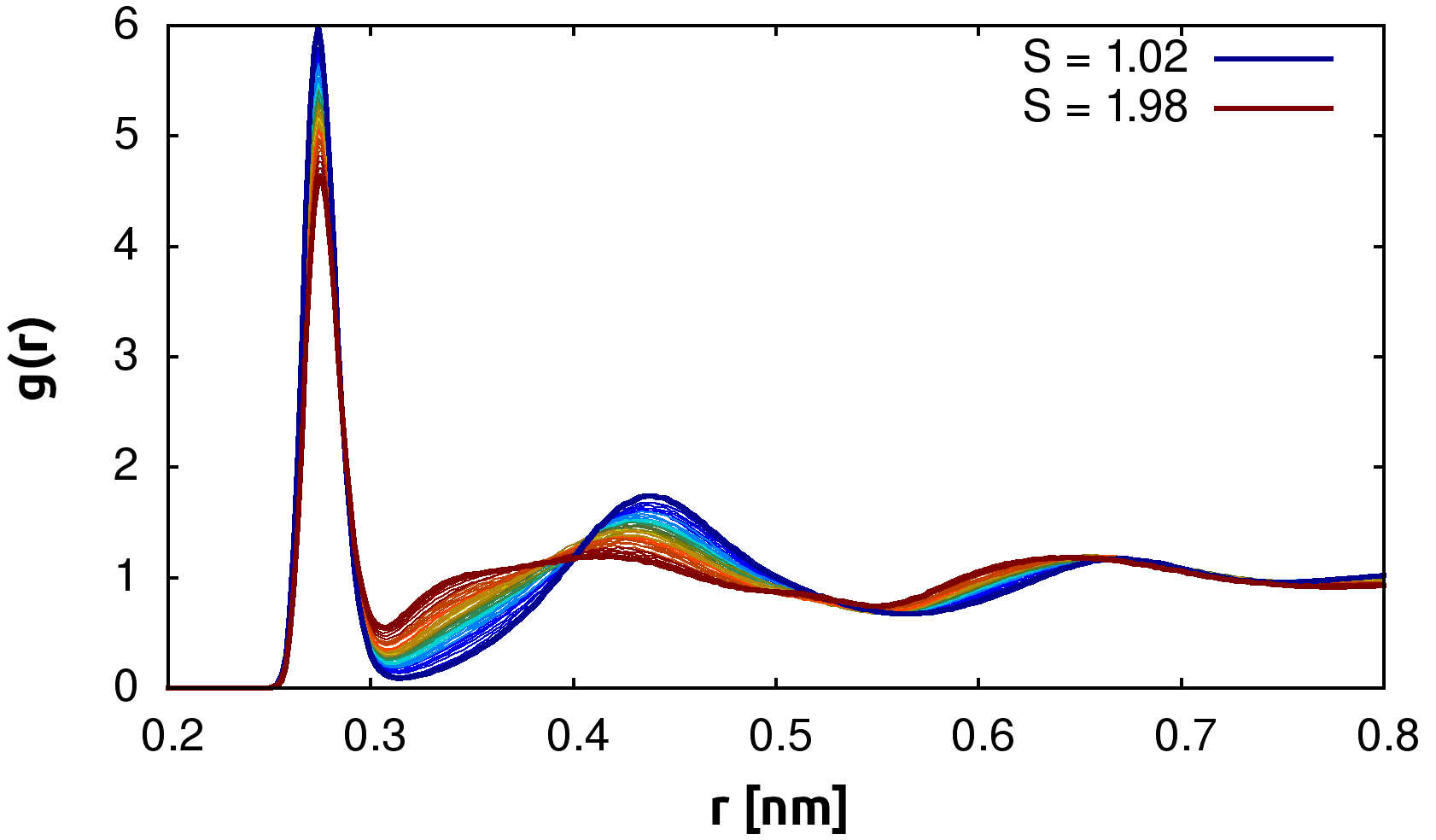}
    \caption{}
    \label{fig:rdf_155K-2250b}
  \end{subfigure}
  \caption{
    Radial distribution function $g(r)$ of all the umbrella sampling windows, discarding the first half of the simulation as equilibration, at 170 K and 2 kbar (\ref{fig:rdf_170K-2kb}), and 155 K and 2.25 kbar (\ref{fig:rdf_155K-2250b}). A continuous transition can be observed from the low (blue) to the high density states (red). Similar results hold for all the other $(P, T)$ conditions addressed in this study.
  }
  \label{fig:rdf_ldl_hdl}
\end{figure*}

\begin{figure*}[!ht]
    \centering
    \includegraphics[width=0.9\linewidth]{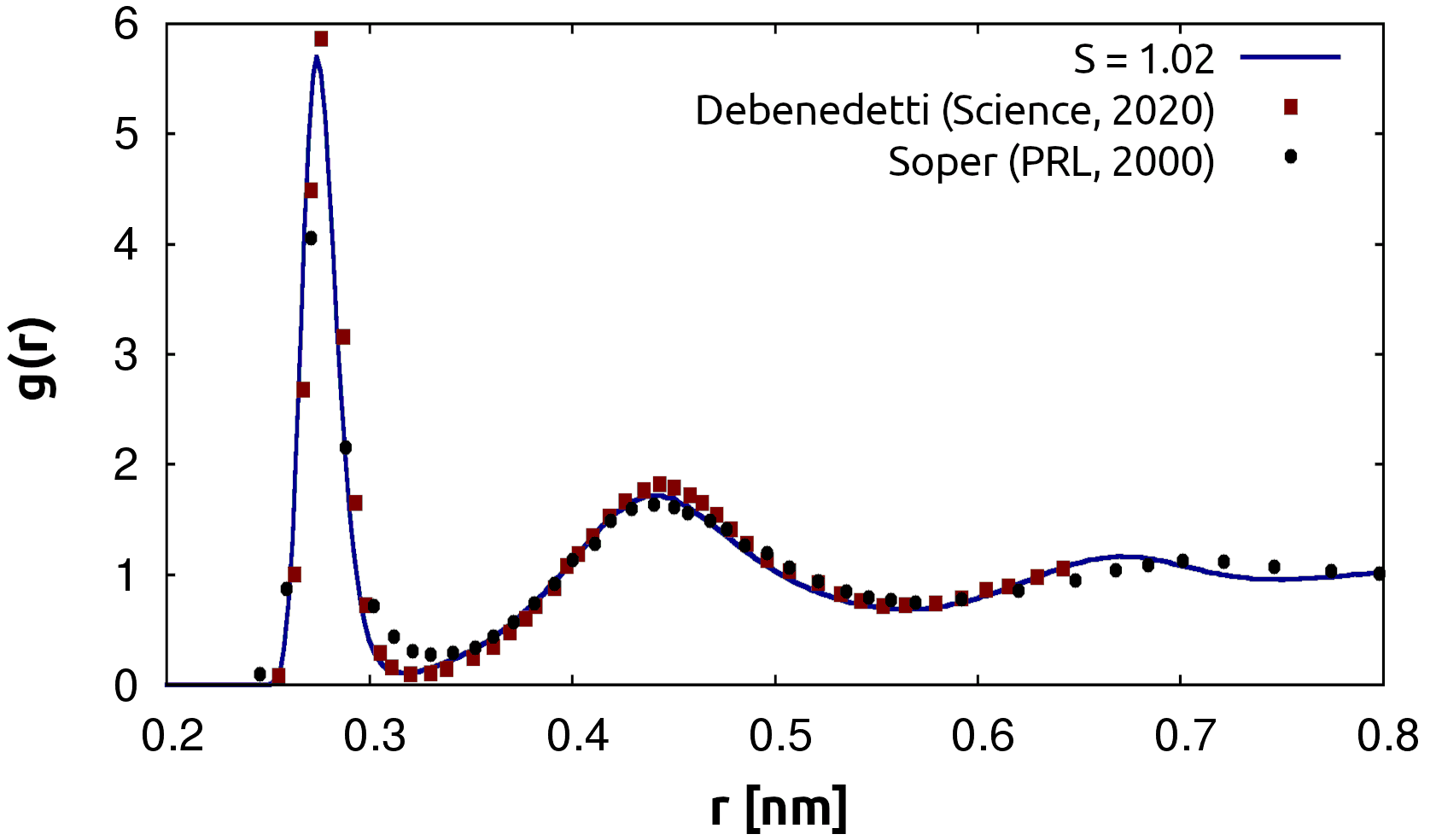}
    \caption{
      Comparison between the radial distribution function $g(r)$ of the low-density state in our work (line), at 170 K and 2 kbar with 800 TIP4P/2005 water molecules, with the one of Ref.~\cite{debenedetti-science-20} (squares), obtained at 177 K and 1.75 kbar with 300 TIP4P/2005 water molecules, and the one of Ref.~\cite{soper-prl-00-ldl-hdl-struct} (dots), obtained with empirical potential structure refinement based on the structure factor measured experimentally at 268 K with neutron diffraction.
      The data for the latter two $g(r)$ are extracted graphically from  Ref~\cite{soper-prl-00-ldl-hdl-struct} and ~\cite{debenedetti-science-20}.
    }
    \label{fig:rdf_ldl_ours_debe_soper}
\end{figure*}

\begin{figure*}[!ht]
    \centering
    \includegraphics[width=0.9\linewidth]{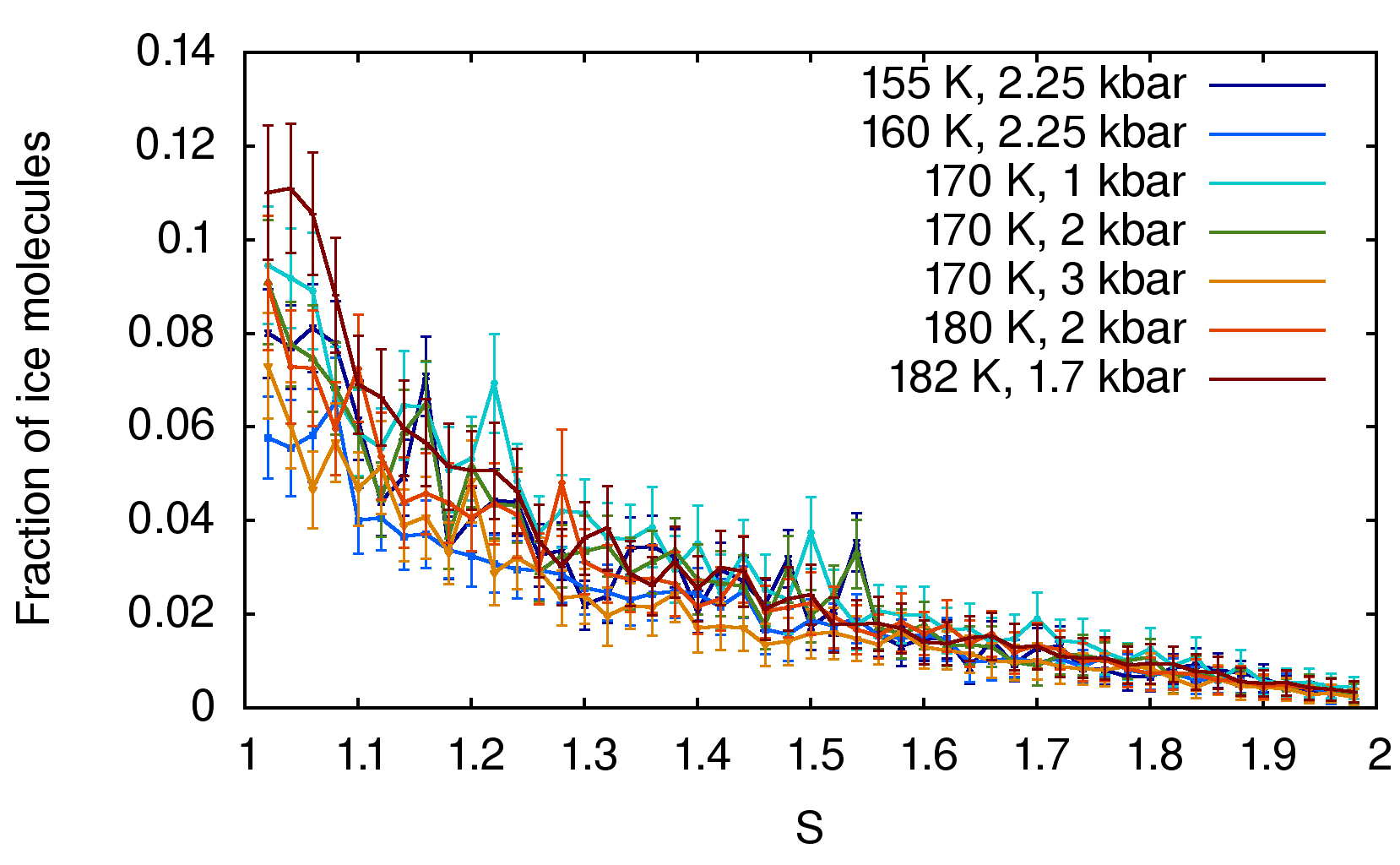}
    \caption{
      Fraction of ice molecules in each umbrella sampling window at different $(P,T)$ conditions, discarding the first half of each trajectory as equilibration. Molecules are classified with the Chill+ algorithm~\cite{nguyen2015identification}: we plot the sum of cubic, hexagonal and interfacial ice molecules, divided by the total number of water molecules.
      $S\sim 1.1$ correspond to low density and $S\sim 1.9$ correspond to high density.
      See Table~\ref{table:ice_proportion} for the fraction of ice molecules  Boltzmann-averaged over each free-energy profile.
    }
    \label{fig:ices_frac}
\end{figure*}

\begin{table*}
  \centering
  \begin{tabular}{crr}
    $P, T$ & $\langle$ Ice proportion $\rangle$ \\
    \midrule
    155 K, 2.25 kbar &  $2 \pm 2 \%$ \\
    160 K, 2.25 kbar &  $3 \pm 2 \%$ \\
    170 K, 1 kbar    &  $9 \pm 5 \%$ \\
    170 K, 2 kbar    &  $3 \pm 2 \%$ \\
    170 K, 3 kbar    &  $0.6 \pm 0.5 \%$ \\
    180 K, 2 kbar    &  $2 \pm 2 \%$ \\
    182 K, 1.75 kbar &  $5 \pm 4 \%$ \\
    \bottomrule
  \end{tabular}
  \caption{Average proportion of ice molecules in the system, computed from the weighted contribution of every umbrella sampling window with the Boltzmann factor $\mathrm{e}^{-G(S)/{k_{\mathrm{B}}}T}$ to obtain equilibrium averages, discarding the first half of the simulations as equilibration. Molecules are classified with the Chill+ algorithm~\cite{nguyen2015identification}. See Fig.~\ref{fig:ices_frac} to have the detail of ice fraction over the whole range of our variable $S$.}
  \label{table:ice_proportion}
  
\end{table*}


\end{document}